\documentclass[fleqn,usenatbib]{mnras}

\usepackage{newtxtext,newtxmath}
\usepackage[T1]{fontenc}

\DeclareRobustCommand{\VAN}[3]{#2}
\let\VANthebibliography\thebibliography
\def\thebibliography{\DeclareRobustCommand{\VAN}[3]{##3}\VANthebibliography}

\usepackage[pdftex]{graphicx}
\usepackage{enumitem}
\usepackage{mathtools}
\usepackage[version=4]{mhchem}
\usepackage{siunitx}

\DeclarePairedDelimiterX\braket[2]{\langle}{\rangle}{#1 \delimsize\vert #2}
\usepackage{float}
\usepackage{array,multirow,graphicx}
\usepackage[dvipsnames]{xcolor}
\definecolor{mygray}{gray}{0.5}
\def\hide#1{{\color{white}#1}}

\title[Global 21-cm: Ionospheric Effects]{Bayesian Data Analysis for Sky-averaged 21-cm Experiments in the Presence of Ionospheric Effects}

\author[E. Shen et al.]{
Emma Shen,$^{1}$\thanks{E-mail: yhs24@cam.ac.uk}
Dominic Anstey,$^{1}$
Eloy de Lera Acedo$^{1,2}$
and Anastasia Fialkov$^{2,3}$
\\
$^{1}$Cavendish Laboratory, University of Cambridge, Cambridge, CB3 0HE, United Kingdom\\
$^{2}$Kavli Institute for Cosmology, Madingley Road, Cambridge, CB3 0HA, United Kingdom\\
$^{3}$Institute of Astronomy, University of Cambridge, Madingley Road, Cambridge CB3 0HA, United Kingdom
}

\date{Accepted XXX. Received YYY; in original form ZZZ}

\pubyear{2022}

\begin{document}
\label{firstpage}
\pagerange{\pageref{firstpage}--\pageref{lastpage}}
\maketitle

% Abstract of the paper
\begin{abstract}
The ionosphere introduces chromatic distortions on low frequency radio waves, and thus poses a hurdle for 21-cm cosmology. In this paper we introduce time-varying chromatic ionospheric effects on simulated antenna temperature data of a global 21-cm data analysis pipeline, and try to detect the injected global signal. We demonstrate that given turbulent ionospheric conditions, more than 5\% error in our knowledge of the ionospheric parameters could lead to comparatively low evidence and high root-mean-square error (RMSE), suggesting a false or null detection. When using a constant antenna beam for cases that include data at different times, the significance of the detection lowers as the number of time samples increases. It is also shown that for observations that include data at different times, readjusting beam configurations according to the time-varying ionospheric conditions should greatly improve the significance of a detection, yielding higher evidences and lower RMSE, and that it is a necessary procedure for a successful detection when the ionospheric conditions are not ideal.

\end{abstract}

% Select between one and six entries from the list of approved keywords.
% Don't make up new ones.

%%%%%%%%%%%%%%%%%%%%%%%%%%%%%%%%%%%%%%%%%%%%%%%%%%

%%%%%%%%%%%%%%%%% BODY OF PAPER %%%%%%%%%%%%%%%%%%

\section{Introduction}

The global or sky-averaged redshifted 21-cm absorption line of neutral hydrogen (HI) is a probe of the early universe; a direct measurement of the line can give us insights into the transformation of the intergalactic medium. The absorption profile over time is determined by the radio background usually assumed to be the cosmic microwave background (CMB), the thermal state of the gas, Lyman-$\alpha$ radiation emitted by the luminous sources scattering off cold primordial gas  \citep{furl}. The challenge lies in the signal being extremely weak compared with the strong galactic foreground in the targeted frequency band.

Although \citet{edges} reported to have found an absorption profile in the form of a \textit{flattened} Gaussian centred at 78 MHz in the sky-averaged 21-cm spectrum, several works have raised concerns on its unphysical parameters and the non-uniqueness of their solution \citep{hillsnat, singh, bevins, sims}, and \citet {saras3} reported residuals incompatible with the EDGES signals.

Foregrounds including galactic synchrotron radiation, thermal free-free emission from galactic dust, and contribution from extragalactic point sources, emitting in the frequency band 30-240 MHz, together are $\sim$4-5 orders of magnitude larger than the global 21-cm signal in brightness temperature \citep{Reis2020, Reis20211}. Together they are expected to have an amplitude of the order of $\sim$0.1 K, representing some of the main difficulties in achieving statistically significant detection. Thus, better understanding of the spatially and spectrally varying foregrounds is critical to measuring the global 21-cm signal. Although the foregrounds are considered spectrally smooth, while the global 21-cm signal is expected to have peaks and troughs associated with absorption and emission features, the chromaticity introduced by the antenna negates this assumption \citep{anstey}. The antennae exhibit chromaticity when operated in large fractional bandwidths. Because neither chromatic antenna beam nor the sky temperature is smooth as a function of frequency, the convolution of the beam and the sky temperature breaks the assumed smoothness after integration.

The ionosphere introduces extra factors to chromatic mixing by absorption and refraction, and its effects scale approximately as $\nu^{-2}$. Its magnitude can be 2-3 orders larger than the global 21-cm signal within the measurement bandwidth \citep{km, burns}. In this paper, we model the ionospheric layers which cause significant chromatic distortion on the foregrounds by refraction and absorption.

This work is a follow-up of the previous paper \citep{shen21}, in which we studied how the chromatic ionospheric effects distort the global 21-cm signal and how they behave over time. In this paper we focus on studying the impediments the chromatic ionospheric effects introduce to the current global 21-cm experiment data analysis pipeline. It is also part of the Radio Experiment for the Analysis of Cosmic Hydrogen, or REACH \citep{reachh}, a novel experiment for the detection of the global 21-cm line from neutral hydrogen red-shifted to a few hundreds MHz due to the expansion of the universe. It is based on state of the art Bayesian analysis informed by physically rooted models of the cosmic signal, foregrounds and the instrument itself. Different cases are compared using Bayesian log evidence and RMSE of the detected signal. We assume that the antenna beam pattern is distorted by the chromatic ionospheric effects and thus it changes with time depending on the ionospheric condition. The antenna temperature data subjected to the time-varying chromatic ionospheric effects in the 50-200 MHz band is analysed using the current REACH data analysis pipeline \citep{anstey}, in which the antenna beam used to fit the foregrounds too is modified by the simulated ionospheric effects using the two-layered ionospheric model. The foregrounds are modelled using one constant beam for data collected at different times and then time-varying beams implemented through the multiple time-steps tested in the simulation. We also investigate the level of error the data analysis pipeline is able to afford, in terms of our knowledge of the parameters we use to model the ionosphere. To find the error threshold, we apply fixed amounts of deviation to the modelled ionospheric parameters when generating the antenna temperature data. 

This paper is ordered as follows. In section \ref{sec2} we briefly describe the ionospheric model we adopt, and how it changes the observation data. Section \ref{sec3} introduces the basis of the data analysis pipeline used in this paper as well as in REACH. In section \ref{sec4}, we show the results of the analyses done by the pipeline on the data including the ionospheric effects, and section \ref{sec5} concludes the paper.

\section{Modelling Chromatic Ionospheric Effects}
\label{sec2}
\subsection{Refraction and Absorption}
We model the F-layer and the D-layer of the ionosphere \citep{bly,nicole,theight,evans1968}, which could cause significant chromatic distortion on the foregrounds by refraction and absorption respectively \citep{km}. Both the F-layer and the D-layer of the ionosphere are assumed to be homogeneous in the model, having no spatial variation across the azimuth and height. The parameters used to model the ionospheric effects are electron density $N_\mathrm{e}$, electron collision frequency $\nu_\mathrm{c}$, layer height, and layer width. The effective beam subject to both the F-layer refraction and the D-layer absorption is
\begin{equation}
\begin{aligned}
\hat{B}(\nu, \theta, \phi) = B(\nu, \theta + \delta\theta, \phi) \mathcal{L}(\nu, \theta + \delta\theta),
\end{aligned}
\label{eqbeam}
\end{equation}
where $\nu$ is the frequency, $\theta$ is the incoming angle, $\delta \theta$ is the refraction angle, $\theta + \delta \theta$ is the effective angle after refraction, and $\mathcal{L}$ is the loss factor due to absorption. One can then calculate the simulated antenna temperature $T_\mathrm{A}$ by the following integral:
\begin{equation}
\begin{aligned}
T_\mathrm{A}(\nu) = \int ^{2\pi}_{0}d\phi\int ^{\pi/2}_{0}d\theta \sin{(\theta)}T_\mathrm{f}(\nu,\theta,\phi)\hat{B}(\nu,\theta,\phi),
\end{aligned}
\label{anttemp}
\end{equation}
where $T_\mathrm{f}$ is the sky temperature at a given time. If we include time dependence in the equation,
eq. (\ref{anttemp}) becomes 
\begin{equation}
\begin{aligned}
T_\mathrm{A}(\nu,t) = \int ^{2\pi}_{0}d\phi\int ^{\pi/2}_{0}d\theta \sin{(\theta)}T_\mathrm{f}(\nu,\theta,\phi,t)\hat{B}(\nu,\theta,\phi,t).
\end{aligned}
\label{anttemptime}
\end{equation}

\subsection{Time-varying Ionospheric Effects}
The real ionosphere varies with time, and the chromatic ionospheric effects on the foregrounds do not cancel out over time. Hence, we model the ionosphere layers based on actual observation data \citep{shen21}. The data are collected from Lowell GIRO Data Center at station Louisvale, South Africa (\url{https://ulcar.uml.edu/DIDBase/}). Data of the parameters used to characterise the D-layer are mostly absent during night time due to low electron density. Therefore, in our model, the D-layer parameters are scaled from the available data of the F-layer parameters based on the fiducial theoretical values.

\section{Data Analysis Pipeline}
\label{sec3}
The REACH data analysis pipeline \citep{anstey} uses \texttt{PolyChord}, a nested sampling algorithm based on Bayesian inference \citep{hand5b}. It adopts a physically motivated method of modelling the foregrounds of 21-cm experiments in order to fit the systematic chromatic distortions as part of the foregrounds.

\subsection{Bayesian Inference}
Bayesian inference is useful in performing parameter estimation and model comparison. Algorithms based on Markov-Chain Monte-Carlo (MCMC) methods, such as the Metropolis Hastings (MH) algorithm \citep{sidd} and its more sophisticated variants \citep{Mackey2013}, are used to achieve parameter estimation. Model comparison requires calculating evidence, which is the integration of likelihood over the prior density. 

By Bayes' theorem, one can derive the equation
\begin{equation}
\begin{aligned}
\mathrm{P}(\theta_\mathcal{M}|\mathcal{D},\mathcal{M}) &= \frac{\mathrm{P}(\mathcal{D}|\theta_\mathcal{M},\mathcal{M})}{\mathrm{P}(\theta_\mathcal{D}|\mathcal{M})}, \quad \mathrm{or} \quad
\mathcal{P} = \frac{\mathcal{L}\pi}{\mathcal{Z}},
\end{aligned}
\label{eqbayes}
\end{equation}
 where $\mathcal{P}$ is the posterior distribution. $\mathcal{L}$ is the likelihood, which is the probability of the data given a model $\mathcal{M}$ and  the set of parameters $\theta_\mathcal{M}$ describing the model. $\mathcal{\pi}$ is the prior distribution of the parameters, and $\mathcal{Z}$ is the Bayesian evidence or marginal likelihood:
\begin{equation}
\begin{aligned}
\mathcal{Z} = \int \mathrm{P}(\mathcal{D}|\theta_\mathcal{M},\mathcal{M})\mathrm{P}(\theta_\mathcal{D}|\mathcal{M})d\theta_\mathcal{M} = \int \mathcal{L}\pi d\theta_\mathcal{M}.
\end{aligned}
\label{eqev}
\end{equation}
In this paper, we calculate the value of likelihood $\mathcal{L}$ to estimate the confidence of the signal detection in the data analysis pipeline.

\subsection{Foreground Modelling}

A sky model $T_{\mathrm{sky}}(\nu, \theta, \phi)$ is generated across the observing band by dividing the sky into $N$ regions of similar spectral indices and scaling a base map assuming a distinct uniform spectral index in each region. This sky map is then convolved with a antenna beam model $B(\nu, \theta, \phi)$ to produce a model of foregrounds and chromaticity parameterised by the spectral indices of the regions. In this work, number of regions is $25$. An approximate sky map and an antenna directivity pattern generated from physically motivated parameters for each observing frequency are convolved by the equation
\begin{equation}
\begin{aligned}
T_{\mathrm{D}} = \frac{1}{4\pi} \int^{4\pi} _0 B(\nu, \theta, \phi) \int^{t_{\mathrm{end}}} _{t_{\mathrm{start}}} T_{\mathrm{sky}}(\nu, \theta, \phi) dt d\Omega + \hat{\sigma},
\end{aligned}
\label{eqanttemp}
\end{equation}
where $\hat{\sigma}$ is uncorrelated Gaussian noise, to produce a parameterised model that describes both the foregrounds and the chromatic distortion of the antenna. Then we use the Bayesian nested sampling algorithm \texttt{PolyChord} to fit this foreground model to the data with a parameterised 21-cm signal model \citep{anstey} .

\subsection{Likelihood of the Local Sidereal Times (LST) Model}

The model adopted in this work is the local sidereal times (LST) model, in which the data are not integrated over the observing time \citep{domlst}. The general form of likelihood for the LST model is:
\begin{equation}
\begin{aligned}
\log{\mathcal{L}} = \sum_i\sum_j\left[-\frac{1}{2}\log(2\pi\theta_{\sigma}) - \frac{1}{2}\left(\frac{T_{\mathrm{D}i,j}-T_{\mathrm{M}i,j}(\theta_\mathrm{F},\theta_\mathrm{S})}{\theta_{\sigma}}\right)^2\right],
\end{aligned}
\label{eqlike}
\end{equation}
where \textit{i} indicates frequency bins and \textit{j} indicates time bins. $T_D$ is the observation data and $T_M$ is the parameterised model. In this equation, $\theta_\mathrm{F}, \theta_\mathrm{S}, \theta_\sigma$ are foreground, signal, and noise parameters respectively. $T_{\mathrm{M}i,j}(\theta_\mathrm{F},\theta_\mathrm{S})$ can be split into foreground $T_\mathrm{F}$ and signal $T_\mathrm{S}$ components. The global 21-cm signal has no time dependencies, meaning eq. (\ref{eqlike}) can be written as
\begin{equation}
\begin{aligned}
\log{\mathcal{L}} = -\frac{1}{2}\sum_i&\sum_j\log(2\pi\theta_{\sigma}) \\ &- \frac{1}{2}\sum_i\sum_j\left(\frac{T_{\mathrm{D}i,j}-T_{\mathrm{F}i,j}(\theta_\mathrm{F})-T_{\mathrm{S}i}(\theta_\mathrm{S})}{\theta_{\sigma}}\right)^2.
\end{aligned}
\label{eqlike2}
\end{equation}

Detection of the signal that is considered to be significant should show a high difference between the evidences yielded by the foreground model including a Gaussian signal model, $\log\mathcal{Z}_{\mathrm{s}}$, and the one without, $\log\mathcal{Z}_{\mathrm{ns}}$:  $\Delta \log\mathcal{Z} = \log\mathcal{Z}_{\mathrm{s}}- \log\mathcal{Z}_{\mathrm{ns}}$, the subscript \textit{s} and \textit{ns} indicating that the foreground model includes and excludes a signal respectively. Root-mean-square error (RMSE) of the detected signal with respect to the injected signal is also measured to check that the fitted signal is not actually fitting the foregrounds instead of the injected signal. As the values of likelihood and log evidence is calculated by linearly summing over all data points, more time-steps, meaning more data points, could result in bigger values.  

\section{Results}
\label{sec4}
In practice, data are collected chronologically at points with a fixed interval between each during observation time and analysed. In our work, however, there are gaps between the selected time-steps in most of the cases. It is because the ionosphere can suffer significant change even in a short time frame. One of the aims of this work is to analyse how much error we could afford using the REACH data analysis pipeline, and therefore we start with the more ideal cases. With that in mind, we choose time-steps that yield better evidences or signal detection as a group, and do the error analyses from those. This means the time-steps are not necessarily temporally close to each other, and the intervals between the selected time-steps are meaningless in this work. The injected Gaussian signal adopted in the paper has an amplitude of 0.53 K and a standard deviation of 18.7 MHz, centering at 78.1 MHz, which is compatible with the  EDGES \citep{edges} signal in amplitude and centre frequency.

We fit the foreground model using one constant beam for all time-steps or more beams that correspond to the different time-steps, distorted by refraction and absorption due to the ionosphere based on the ionospheric model, to estimate how frequently the beam should be redetermined to achieve a meaningful detection. Fixed errors are introduced to the ionospheric parameters when generating the antenna temperature data, and the data are fitted using the original beam (the beam without applied errors) in the data analysis pipeline to see how much deviation in the parameters the foreground model is able to tolerate.

The location of the antenna used to simulate the data is set in the Karoo radio reserve in South Africa (latitude: \ang{30.7}S, longitude: \ang{21.4}E, height: 25 m). The data for the ionospheric parameters are collected from Lowell GIRO Data Center at station Louisvale, South Africa (\url{https://ulcar.uml.edu/DIDBase/}). The time-steps taken are at the time specified after \ang{00}00'00'' on 1 Jan 2019.

\subsection{Foregrounds Modelled with a Constant Beam}
\label{s41}

\subsubsection{Two time-steps}
We select antenna temperature data at four different time-steps $[t_1, t_2, t_3, t_4]$ that yield reasonably good $\Delta \log\mathcal{Z}$ and relatively low RMSE ($<0.1$K) with that of time-step $t_0$ when analysed using the LST version of the REACH data analysis pipeline, and the foreground signal is modelled using the same beam ($B_{t_0}$) for two different time-steps $[t_0, t_i]$, the log evidences $\log\mathcal{Z}$ of which are shown in Table \ref{tabpap1}. All analyses shown in this work are based on the time-steps listed in Table \ref{tabpap1}.

\begin{figure*}
        \centering
        \begin{tabular}{l}
        \hide{xxxxxxxxxxxxxxxxxxxxxx}\large{$t_0$ \hide{xxxxxxxxxxxxxxxxxxxxxxxxxx}  $t_1$} \\
        \includegraphics[trim={0 0 5cm 0cm},clip,height=5.9cm]{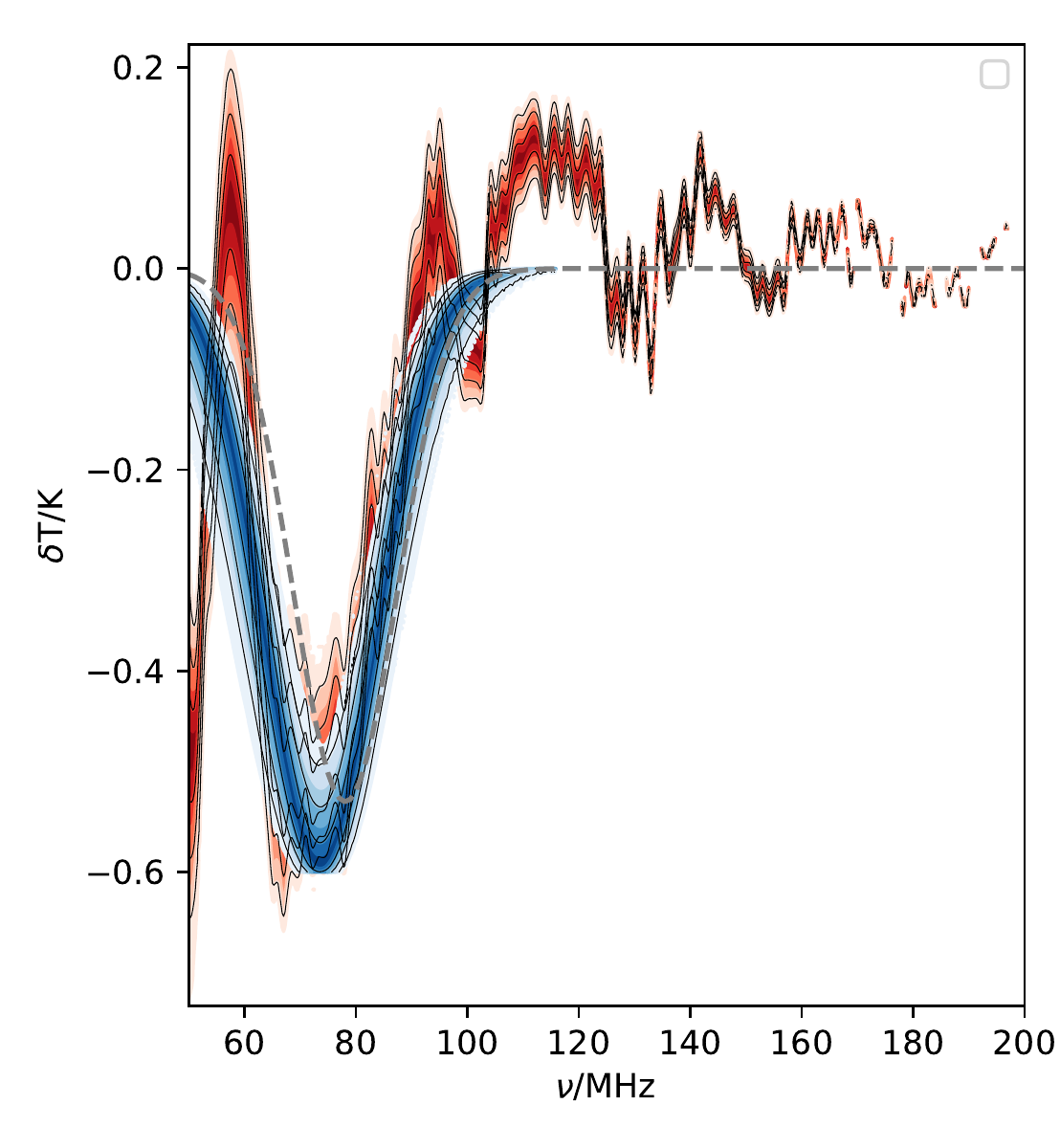} 
        \includegraphics[trim={1cm 0 0.7cm 0},clip,height=5.9cm]{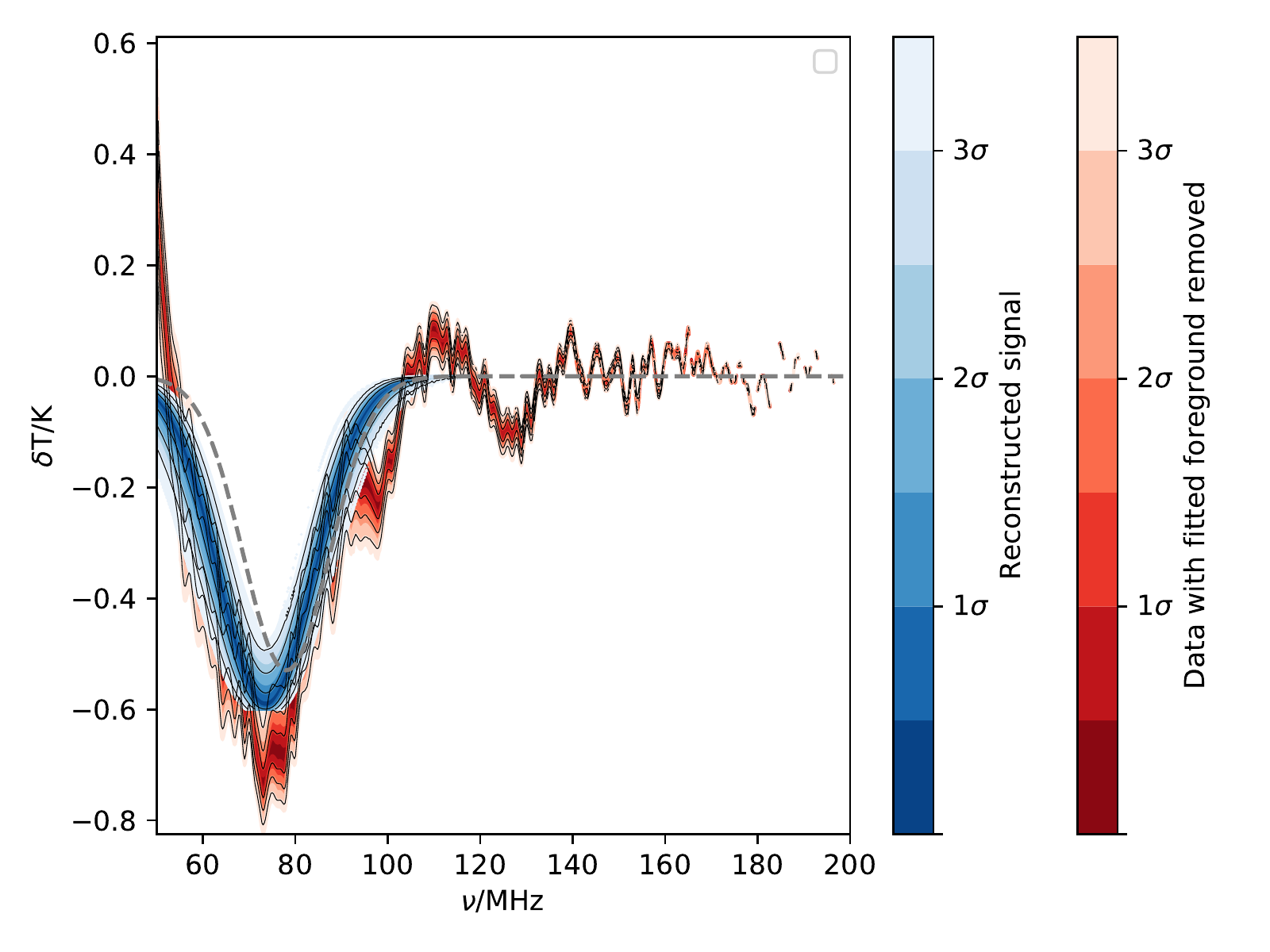}
        \includegraphics[trim={1cm 0 5cm  0},clip,height=5.9cm]{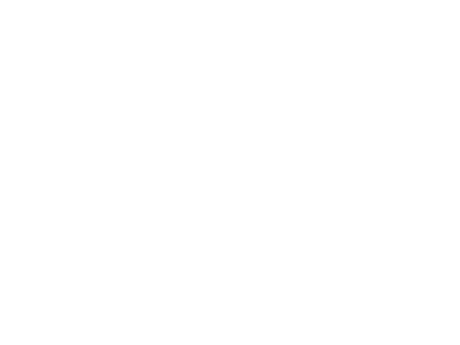} \\ [-15pt]
        \end{tabular}
    \caption{A two time-step case, the foregrounds fitted using one single beam $B_{t_0}$ (case $C_1[t_0,t_1]$ of Table \ref{tabpap1}, the RMSE of the reconstructed signal is 0.066 K and $\Delta\log\mathcal{Z} = 62$). The reconstructed global 21-cm signal is shown in blue, and the data with the fitted foregrounds removed is shown in red. The gray dashed line is the injected signal. The detected signal (reconstructed signal) is significant and the residuals are low for both time-steps.}
    \label{fignb0c1}
\end{figure*}

The settings used to model the foregrounds in the four cases with two time-steps are identical, so the value of the log evidence $\log(\mathcal{Z}_{s})$ should be comparable, as demonstrated by the negative correlation between $\log(\mathcal{Z}_{s})$ and RMSE; a higher $\log(\mathcal{Z}_{s})$ could suggest better signal detection. The fitted signal and residual plots of case $C_1[t_0,t_1]$ are shown in Fig. \ref{fignb0c1}. As one can see, in these cases, the  signal, shown in red, are fitted at significant detection.

\begin{table}
	\centering
	\caption{RMSE \& Log evidence $\log\mathcal{Z}$: the foreground signal is modelled using one single beam ($B_{t_0}$). Case ($C_i$) gives the explored cases. [t] lists all the time-steps analysed in each case. RMSE is the root-mean-square error of the detected signal with respect to the injected signal. $\log\mathcal{Z}_{\mathrm{s}}$ is the evidence yielded by the foreground model including a Gaussian signal model, and $\Delta \log\mathcal{Z}$ is the difference between the evidence yielded by the foreground model including a Gaussian signal model and the one without.}
	\label{tabpap1}
	\begin{tabular}{cccrr} % n columns, alignment for each
		Case & [t] & RMSE (K) & $\log\mathcal{Z}_{\mathrm{s}}$ & $\Delta\log\mathcal{Z}$\\
		\hline

		$C_1$ & $t_0$, $t_1$ & 0.066 & $231.5   \pm 0.3$ & 62.8  \\
		$C_2$ & $t_0$, $t_2$ & 0.035 & $520.2   \pm 0.3$ & 180.4 \\
		$C_3$ & $t_0$, $t_3$ & 0.056 & $93.3    \pm 0.4$ & 161.8 \\
		$C_4$ & $t_0$, $t_4$ & 0.024 & $519.0   \pm 0.3$ & 253.5 \\
		\hline 
		\multicolumn{5}{l}{ $[t_0,t_1,t_2,t_3,t_4]=$ [\ang{21}00', \ang{21}15', \ang{45}15', \ang{46}15', \ang{47}45'] }\\
	\end{tabular}
\end{table}

\subsubsection{Three or more time-steps}
\label{ss412}

\begin{figure*}
        \centering
        \begin{tabular}{l}
        \hide{xxxxxxxxxxxxxxxxxxxxxx}\large{$t_0$ \hide{xxxxxxxxxxxxxxxxxxxxxxxxxx}  $t_1$ \hide{xxxxxxxxxxxxxxxxxxxxxxxxxx}  $t_2$} \\
        \includegraphics[trim={0 0 5cm 0cm},clip,height=5.9cm]{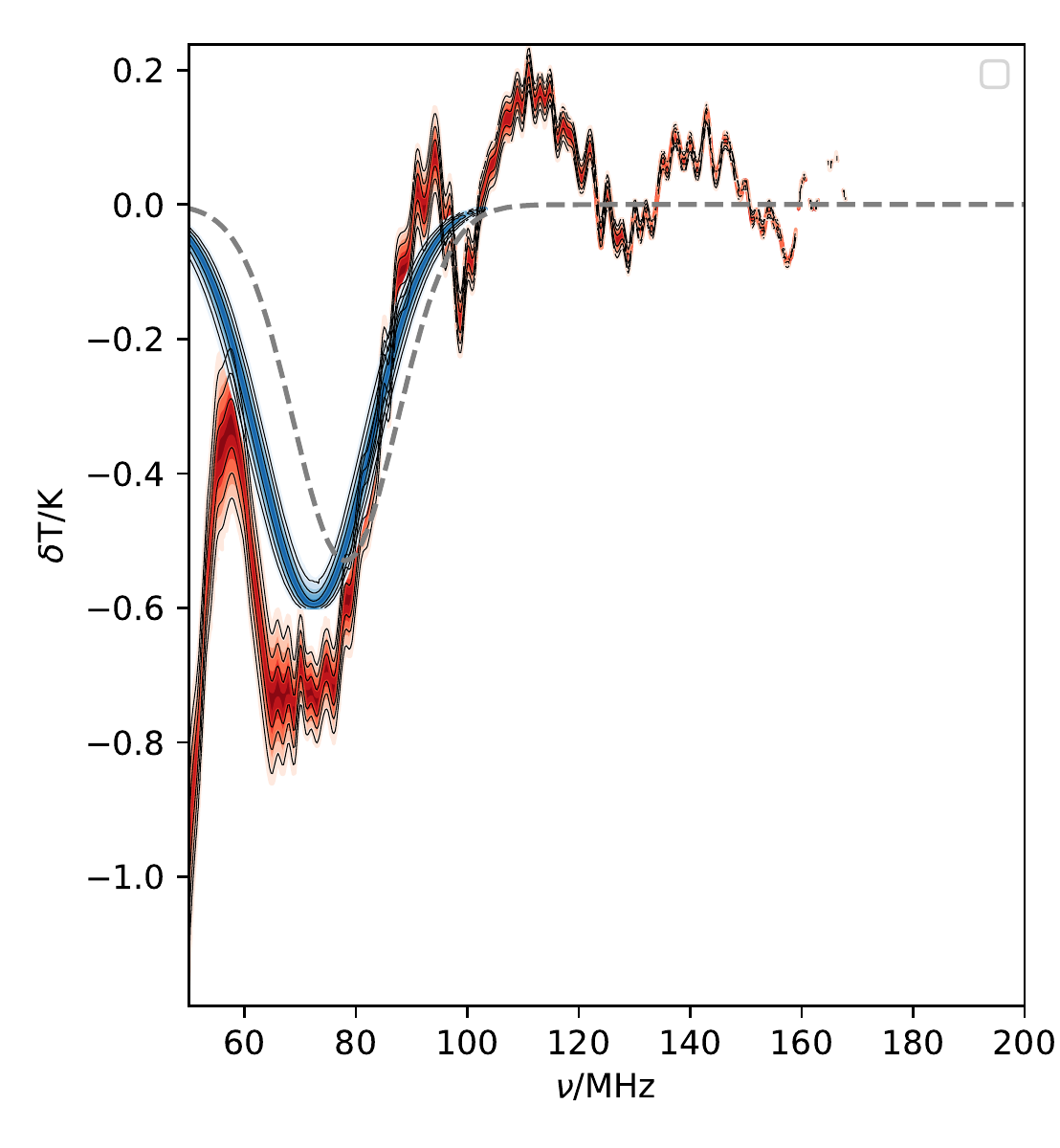}
        \includegraphics[trim={1cm 0 5cm  0},clip,height=5.9cm]{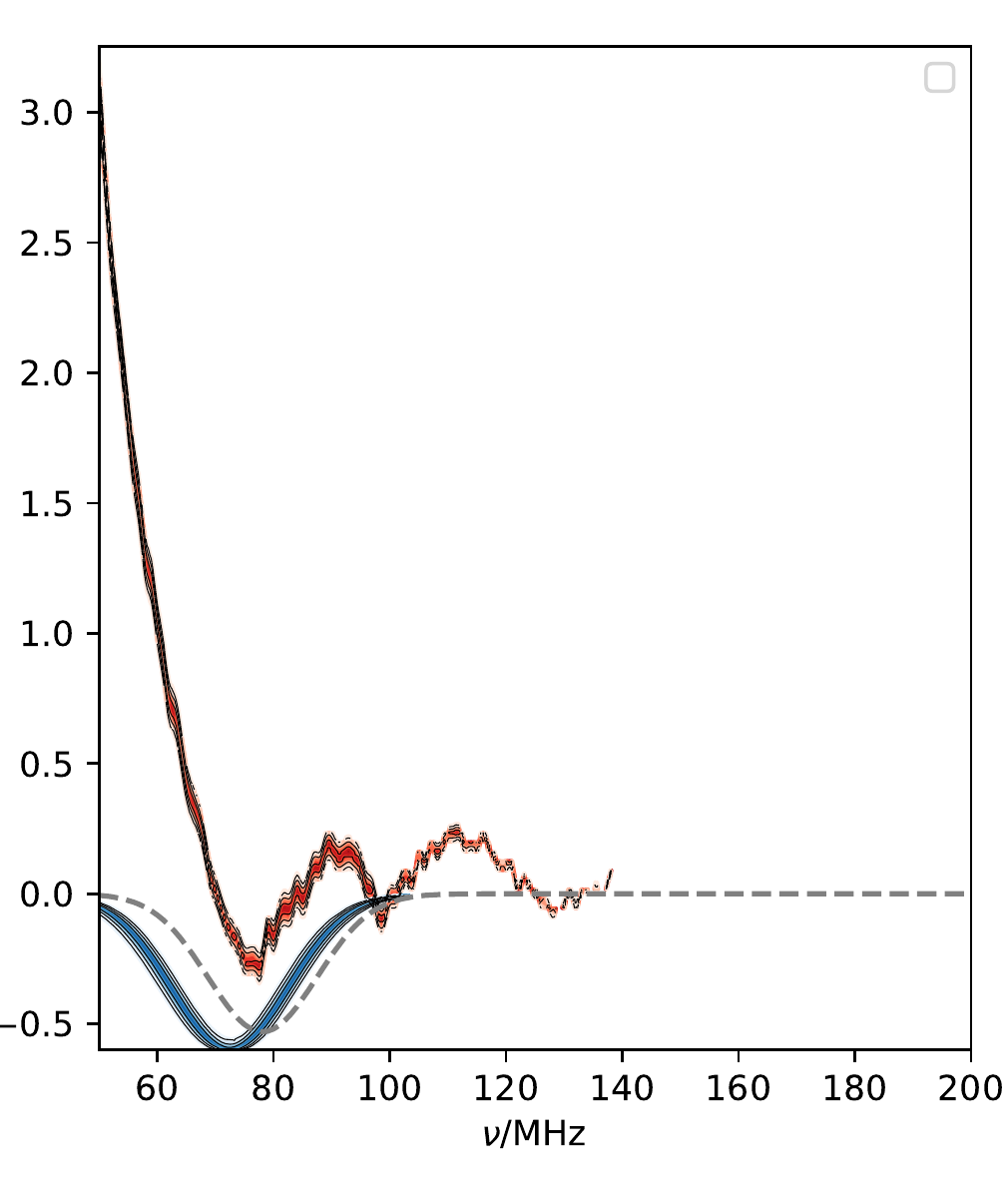}
        \includegraphics[trim={1cm 0 0.7cm 0},clip,height=5.9cm]{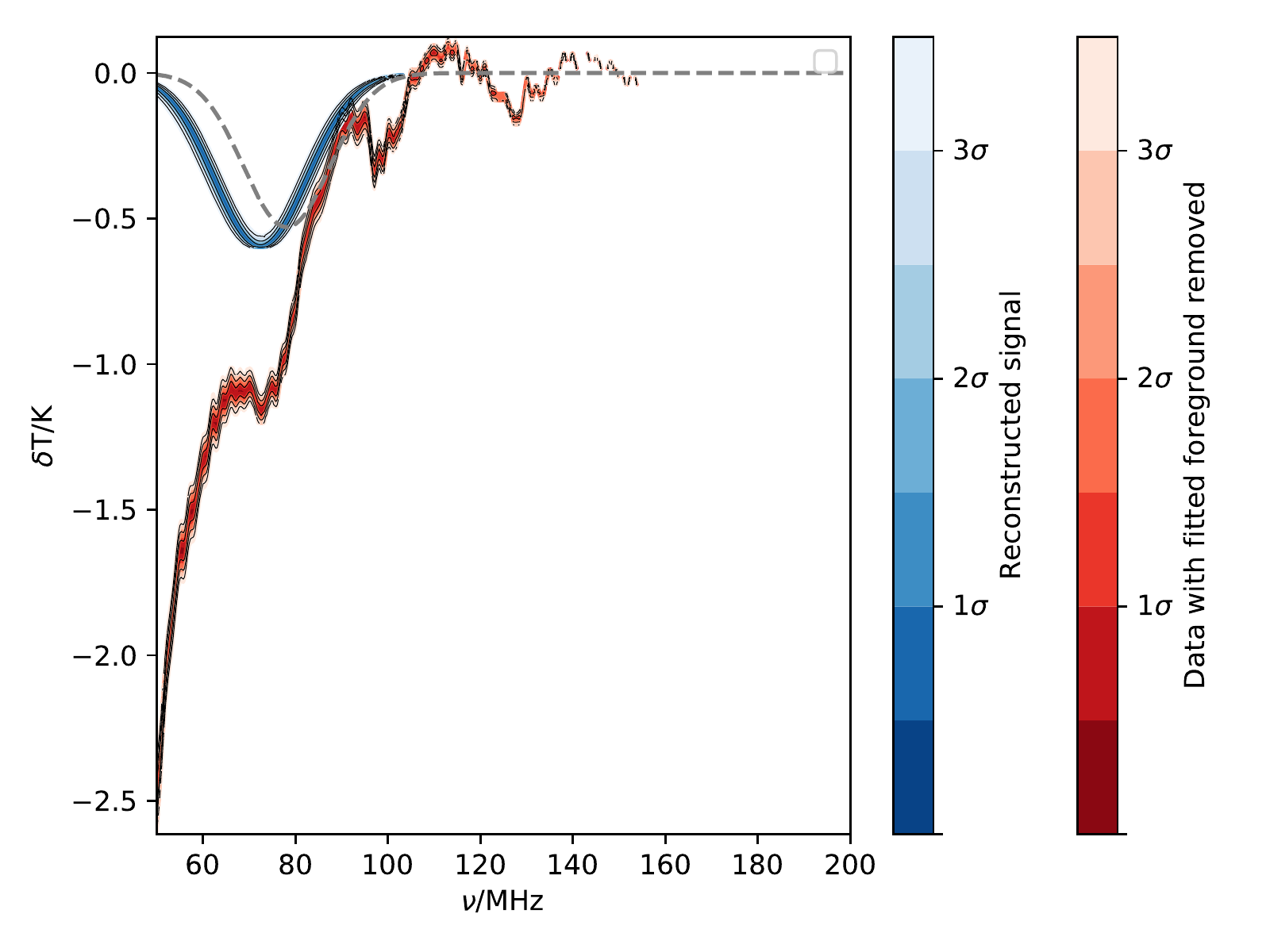}\\[-15pt]
        \end{tabular}
    \caption{A three time-step case, the foregrounds fitted using one single beam $B_{t_0}$ (Case $C_5[t_0,t_1,t_2]$ of Table \ref{tabmb1}, the RMSE of the reconstructed signal is 0.082 K and $\Delta\log\mathcal{Z}=190$). The reconstructed global 21-cm signal is shown in blue, and the data with the fitted foreground removed is shown in red. The gray dashed line is the injected signal. The detected signal (reconstructed signal) is still significant despite its low evidence $\log\mathcal{Z}$ and high residuals in the last two time-steps.}
    \label{fignb1c5}
\end{figure*}

\begin{figure*}
        \centering
        \begin{tabular}{l}
        \hide{xxxxxxxxxxxxxxxxxxxxxx}\large{$t_0$ \hide{xxxxxxxxxxxxxxxxxxxxxxxxxx}  $t_2$ \hide{xxxxxxxxxxxxxxxxxxxxxxxxxx}  $t_4$}\\
        \includegraphics[trim={0 0 5cm 0cm},clip,height=5.9cm]{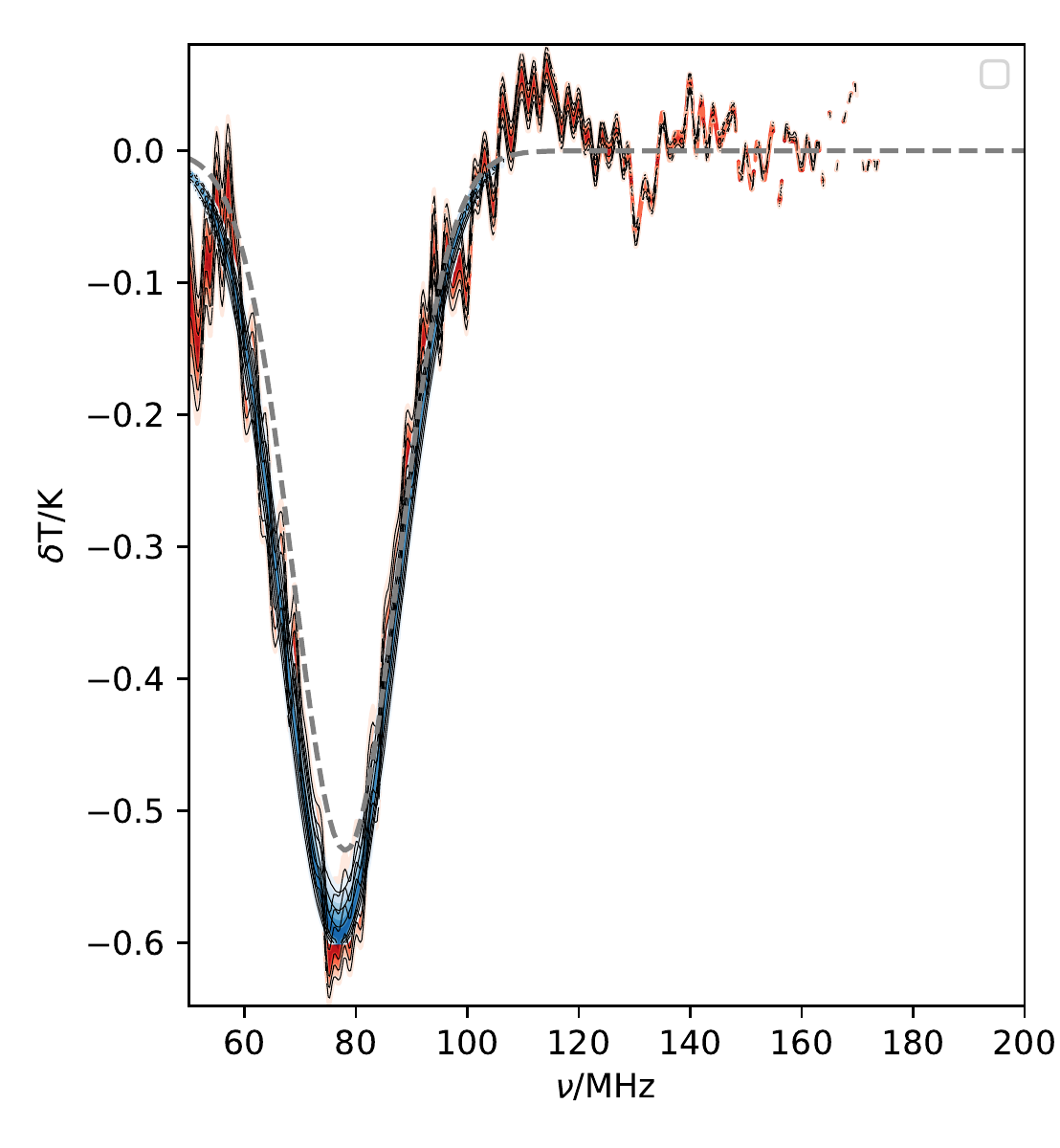}
        \includegraphics[trim={1cm 0 5cm  0},clip,height=5.9cm]{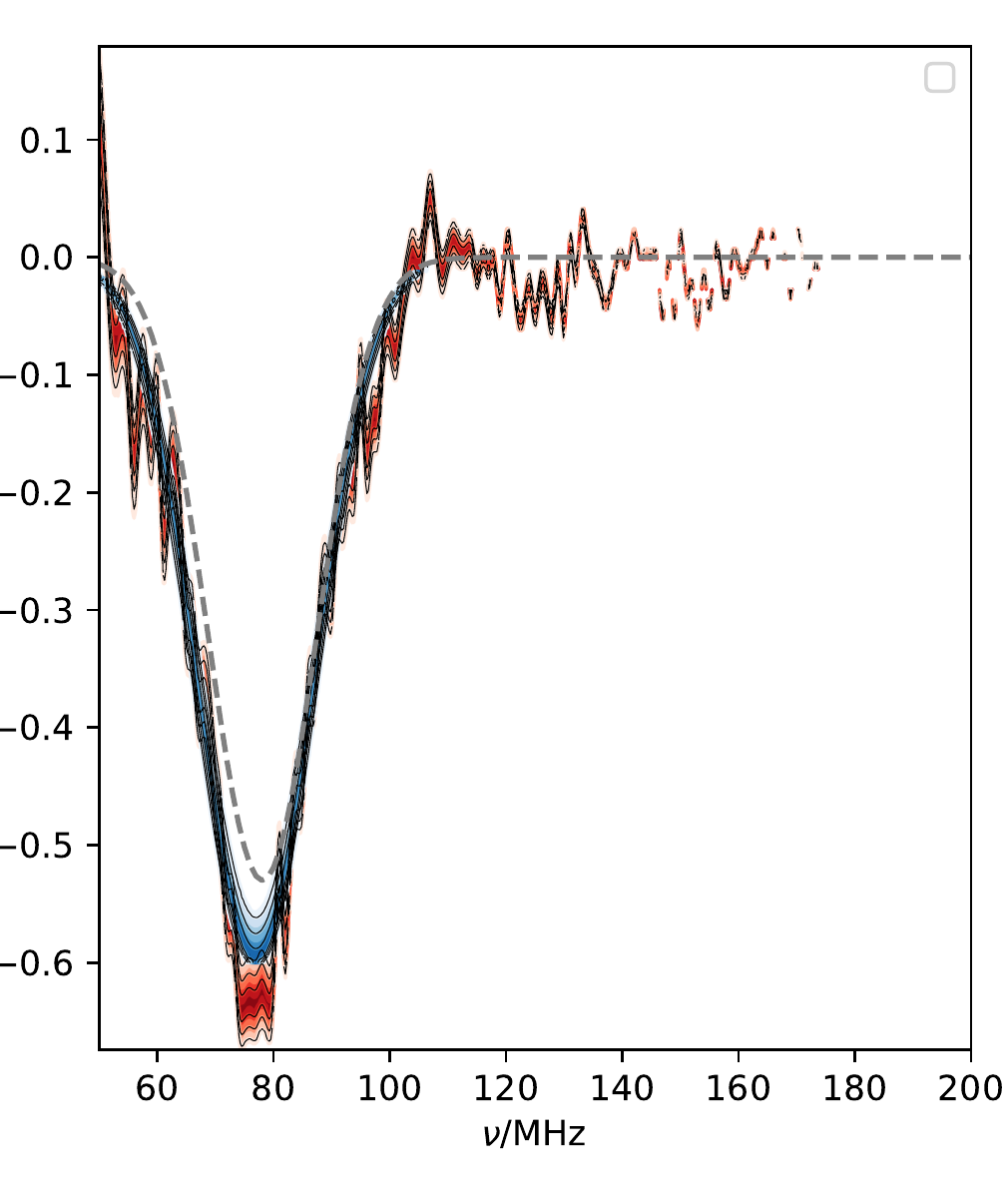}
        \includegraphics[trim={1cm 0 0.7cm 0},clip,height=5.9cm]{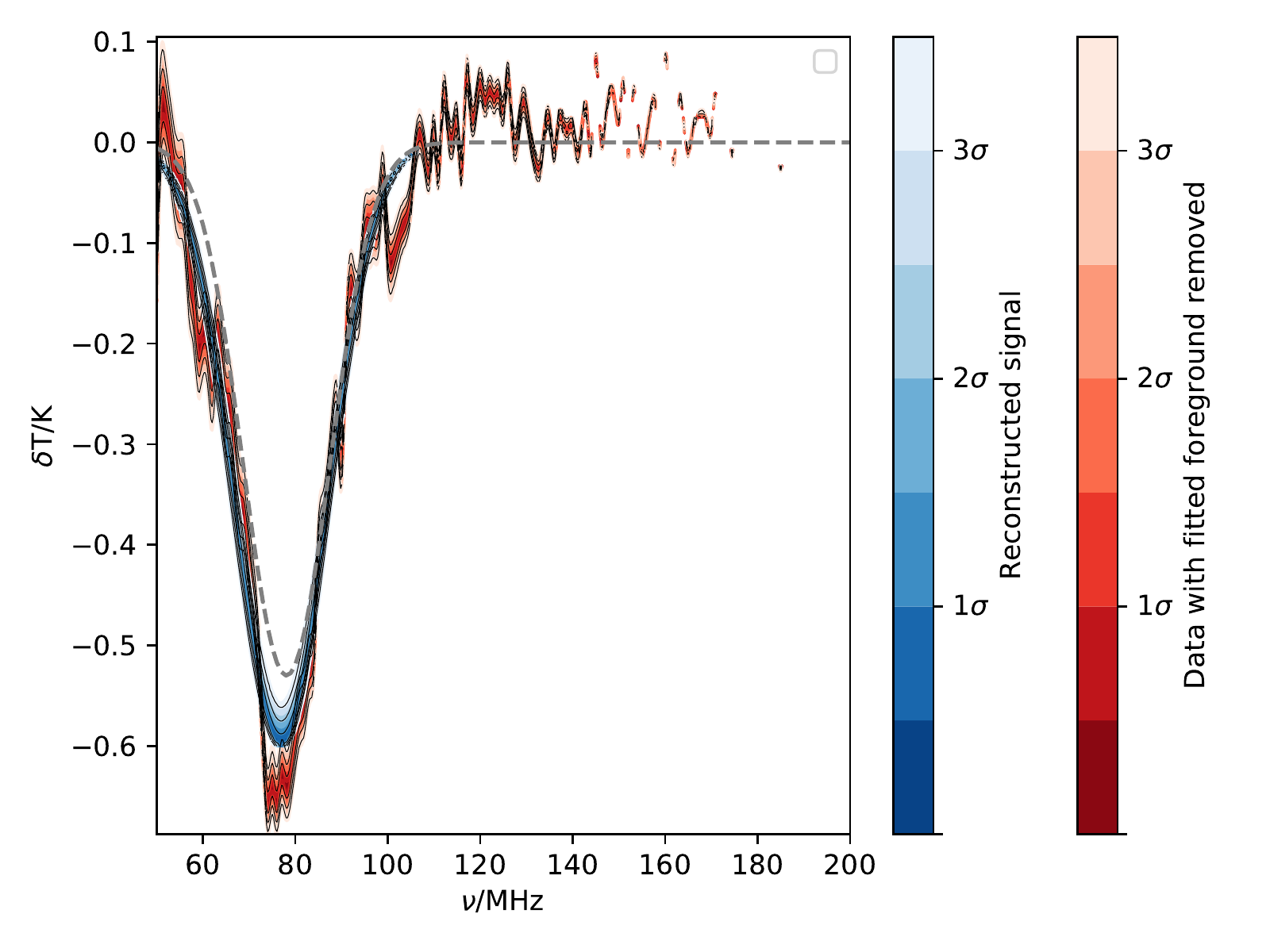}\\[-15pt]
        \end{tabular}
    \caption{A three time-step case, the foregrounds fitted using one single beam $B_{t_0}$ (Case $C_6[t_0,t_2,t_4]$ of Table \ref{tabmb1}, the RMSE of the reconstructed signal is 0.034 K and $\Delta\log\mathcal{Z} = 317$). The reconstructed global 21-cm signal is shown in blue, and the data with the fitted foreground removed is shown in red. The gray dashed line is the injected signal. The detected signal (reconstructed signal) is significant and the residuals are low for all time-steps.}
    \label{fignb1c6}
\end{figure*}

\begin{figure*}
        \centering
        \begin{tabular}{c}
        \large{\hide{xxxx}$t_0$ \hide{xxxxxxxxxxxxxxxxxxxxxxxxxx} $t_1$} \\
        \includegraphics[trim={0 0 5cm 0cm},clip,height=5.9cm]{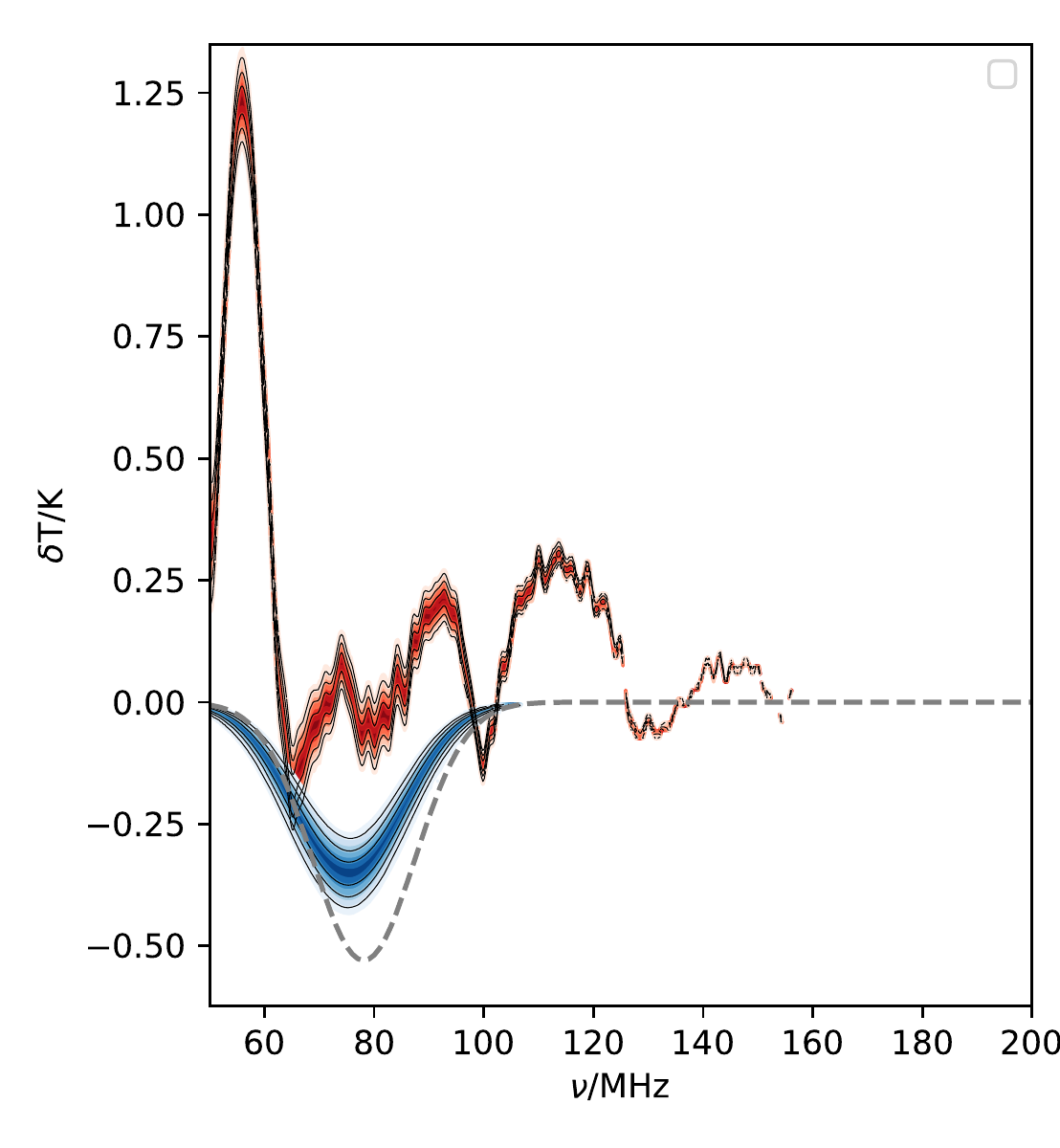}
        \includegraphics[trim={1cm 0 5cm  0},clip,height=5.9cm]{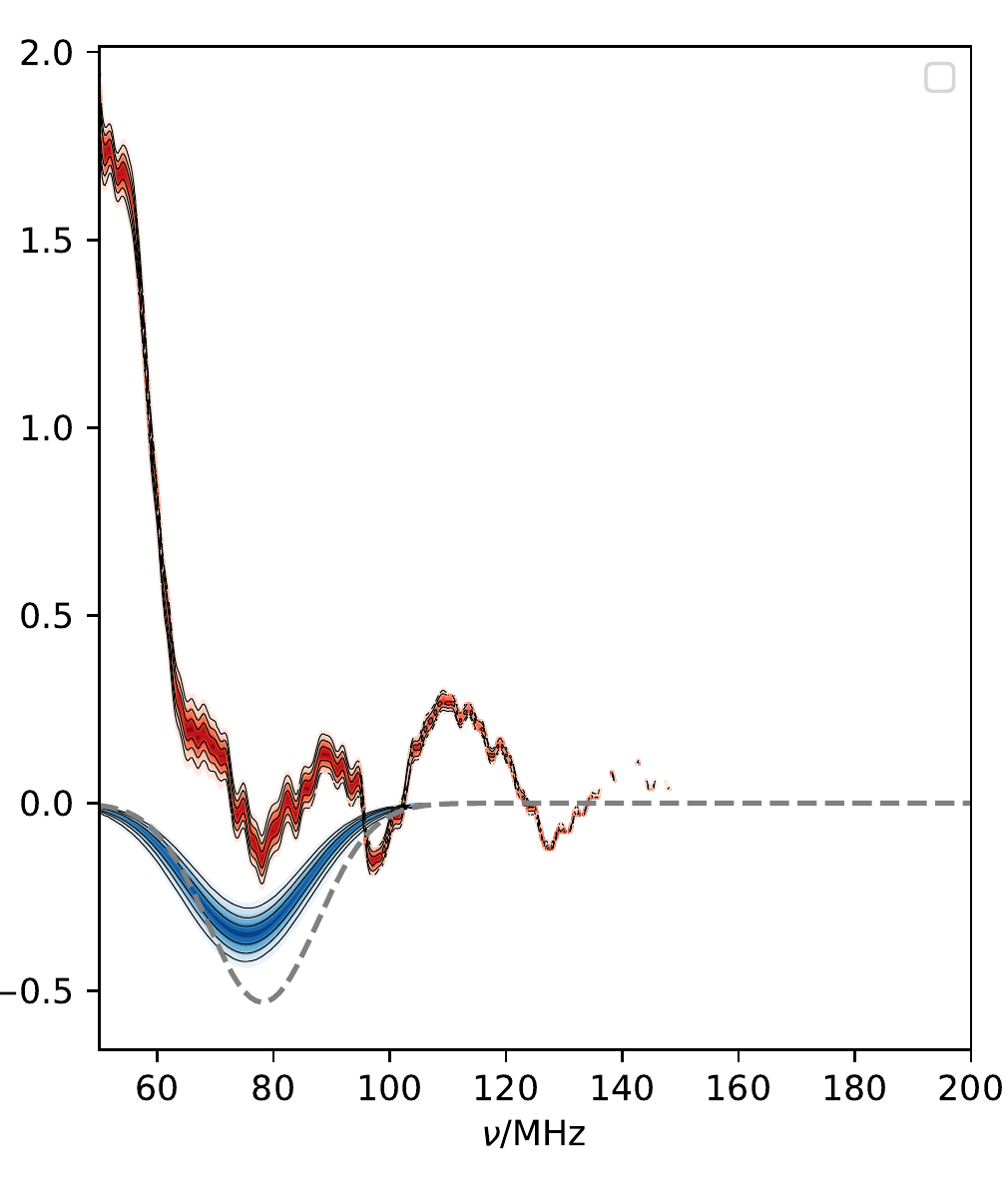}\\
       \large{$t_2$ \hide{xxxxxxxxxxxxxxxxxxxxxxxxxx}  $t_3$ \hide{xxxxxxxxxxxxxxxxxxxxxxxxxx}  $t_4$ \hide{xxxxxxxx}}\\
        \includegraphics[trim={0 0 5cm 0cm},clip,height=5.9cm]{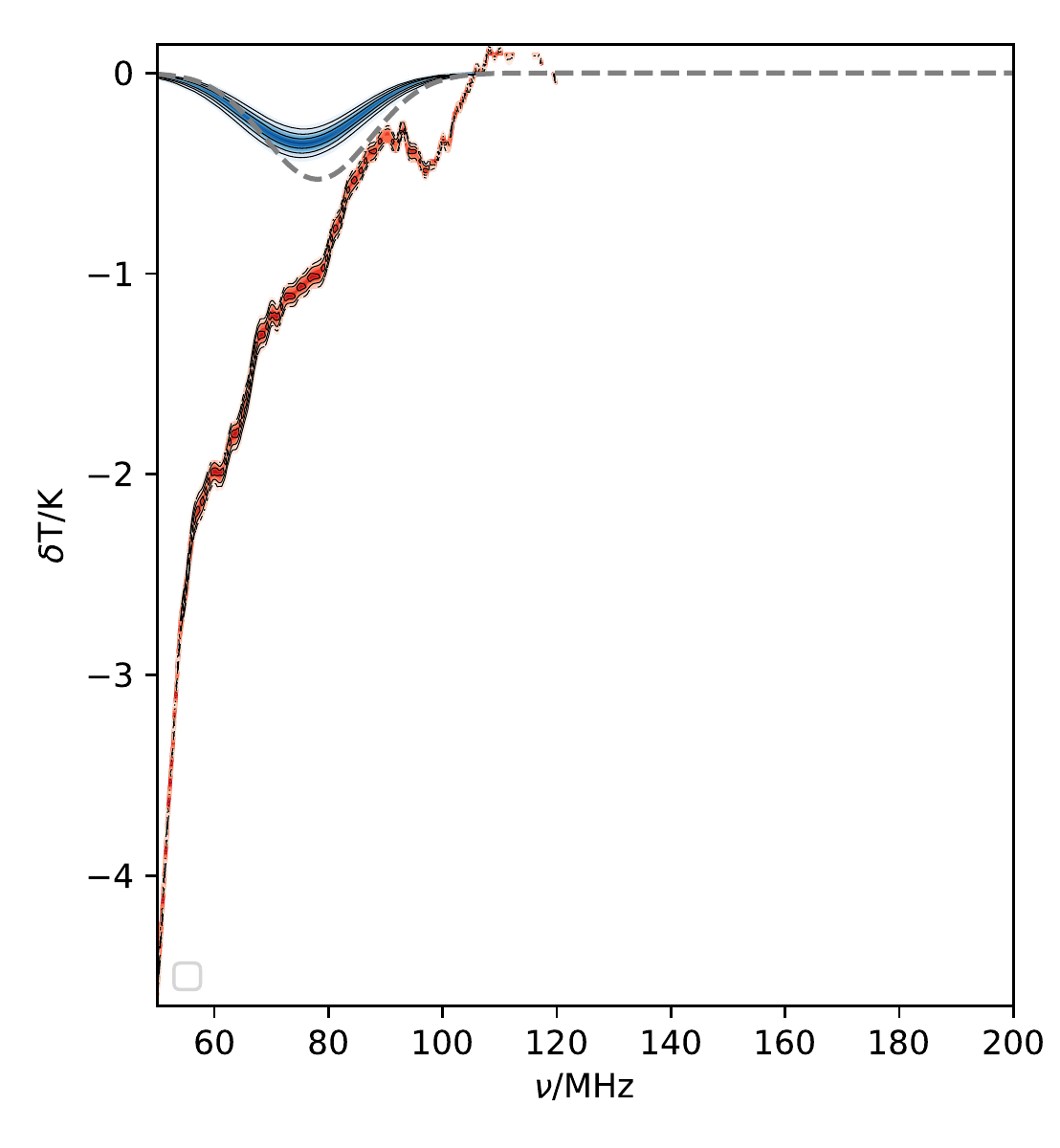}
        \includegraphics[trim={1cm 0 5cm  0},clip,height=5.9cm]{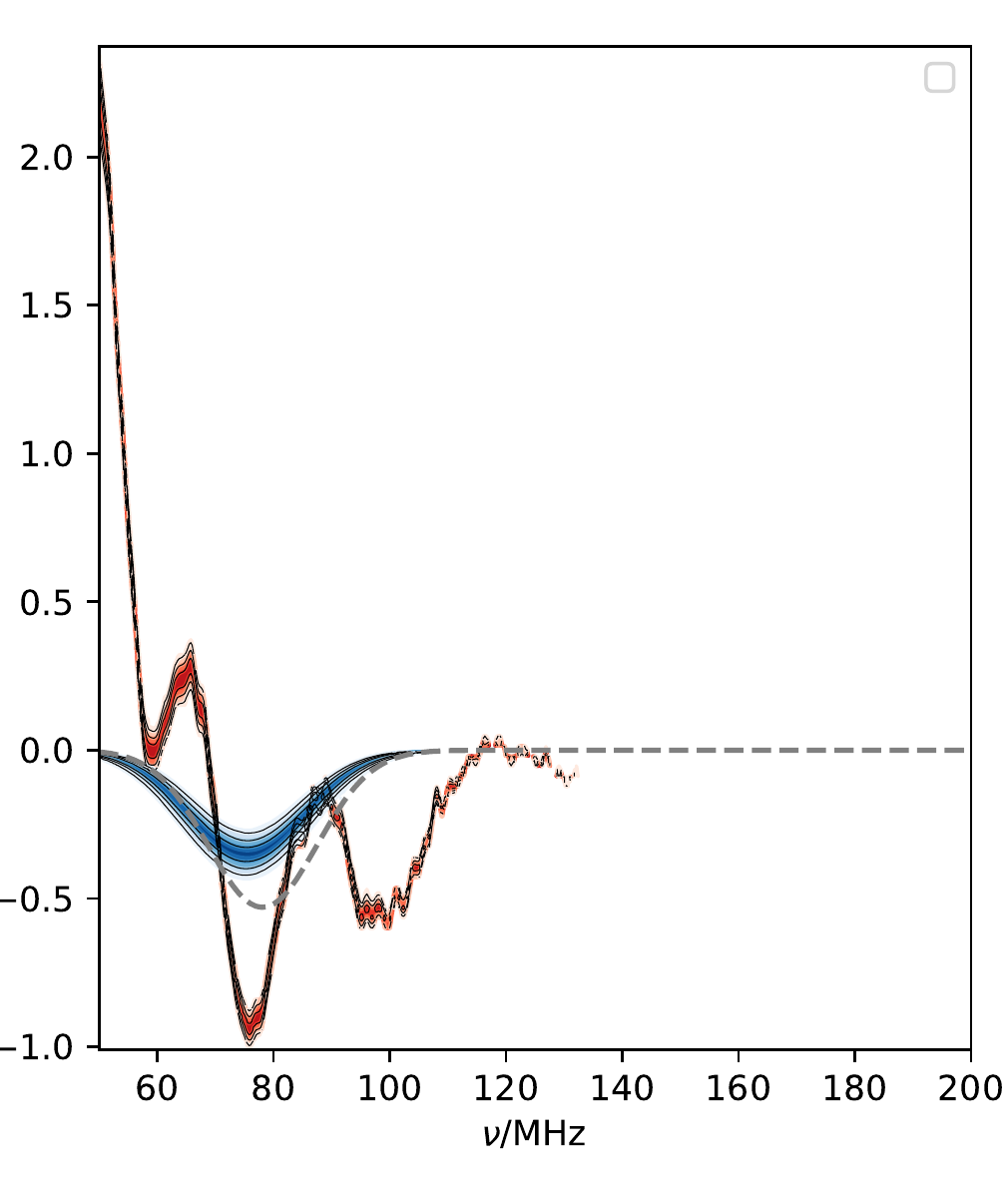}
        \includegraphics[trim={1cm 0 0.7cm 0},clip,height=5.9cm]{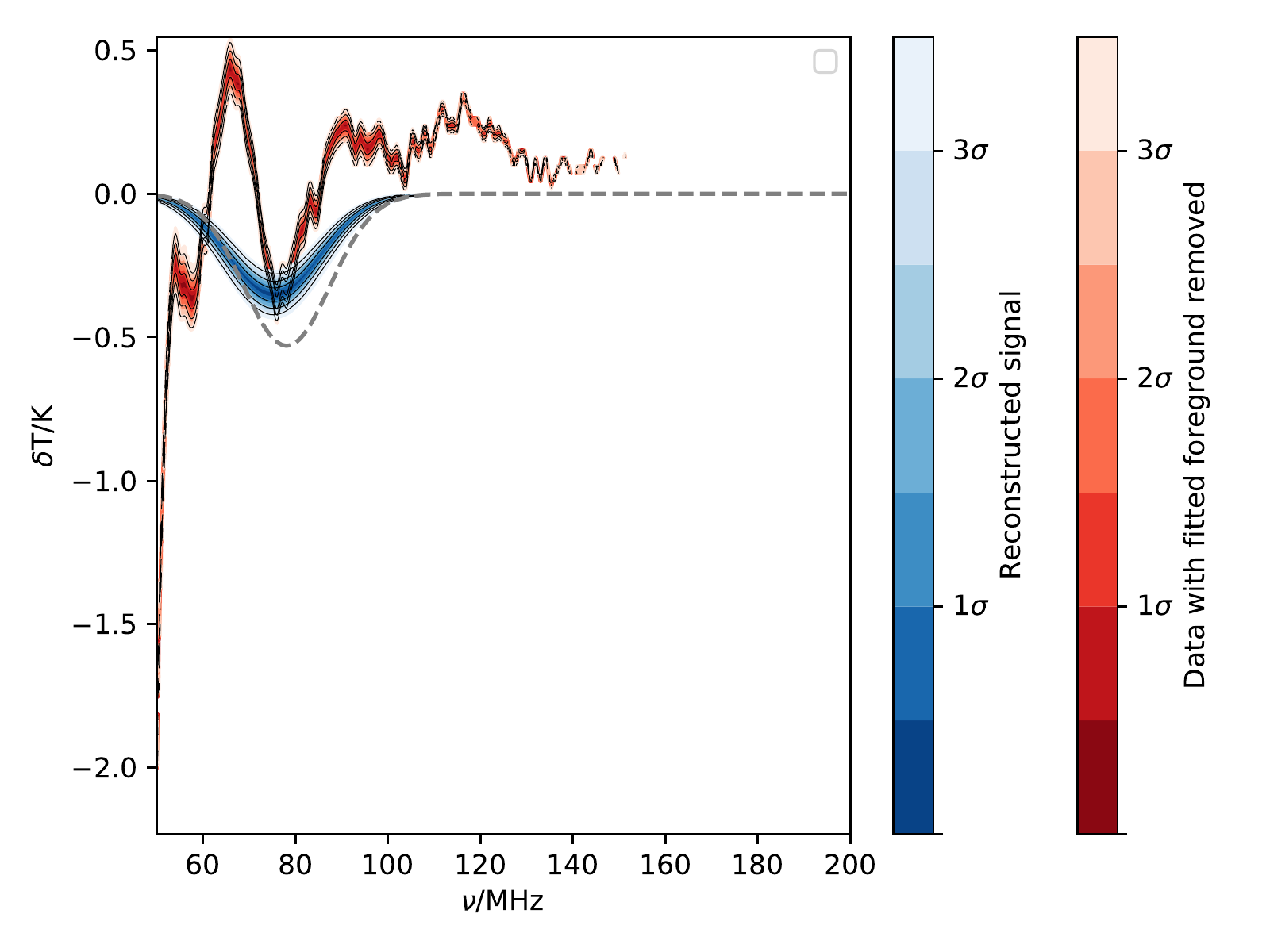} \\[-15pt]
        \end{tabular}
        \caption{A five time-step case, the foregrounds fitted using one single beam $B_{t_0}$ (Case $C_9[t_0,t_1,t_2,t_3,t_4]$ of Table \ref{tabmb1}, the RMSE of the reconstructed signal is 0.063 K and $\Delta\log\mathcal{Z} = 103$). The reconstructed global 21-cm signal is shown in blue, and the data with the fitted foreground removed is shown in red. The gray dashed line is the injected signal. The detected signal (reconstructed signal) is still significant despite its low evidence $\log\mathcal{Z}$ and high residuals.}
    \label{figs191000}
\end{figure*}

Table \ref{tabmb1} shows the evidence of the cases consisting three or more time-steps. It shows that when the foregrounds are modelled using only one beam, all data-sets analysed should have good evidence between themselves, such as $[t_0, t_2]$ and $[t_0, t_4]$ (see Table \ref{tabpap1}), to still yield a good evidence in the cases with more time-steps. For the other cases, one observes that the evidences drop to negative values, but their respective $\Delta \log\mathcal{Z}$ and RMSE still imply significant detection; Fig. \ref{fignb1c5} shows the plot of the fitted signals as well as the residuals of case $C_5[t_0,t_1,t_2]$ at each time-step, and the reconstructed signals in both time-steps are not too different from the injected signal, as implied by the RMSE. The plots of case $C_6[t_0,t_2,t_4]$ are shown in Fig. \ref{fignb1c6}. Despite the negative, low evidences for multiple time-step cases, their RMSE's and reconstructed signals still show significant detection, even in the five time-step case. The plots shown in Fig. \ref{figs191000}. One should note that the time-steps selected for the analyses are the more ideal cases, which might not be the case in actual observations.

\begin{table}
	\centering
	\caption{RMSE \& Log evidence $\log\mathcal{Z}$: the foreground signal is modelled using one single beam  ($B_{t_0}$). Case ($C_i$) gives the explored cases. [t] lists all the time-steps analysed in each case. RMSE is the root-mean-square error of the detected signal with respect to the injected signal. $\log\mathcal{Z}_{\mathrm{s}}$ is the evidence yielded by the foreground model including a Gaussian signal model, and $\Delta \log\mathcal{Z}$ is the difference between the evidence yielded by the foreground model including a Gaussian signal model and the one without.}
	\label{tabmb1}
	\begin{tabular}{crcrr} % n columns, alignment for each
	
		Case & [t] & RMSE (K) & $\log\mathcal{Z}_{\mathrm{s}}$ & $\Delta\log\mathcal{Z}$\\
		\hline

		$C_5$ & $t_0$, $t_1$, $t_2$         & 0.082 & $-4720.2\pm 0.4$    & 194.4\\
		$C_6$ & $t_0$, $t_2$, $t_4$         & 0.034 & $752.6  \pm 0.4$    & 316.1\\
		$C_7$ & $t_0$, $t_1$, $t_2$, $t_4$  & 0.055 & $-4665.1\pm 0.4$    & 196.5\\
		$C_8$ & $t_0$, $t_2$, $t_3$, $t_4$  & 0.083 & $-7308.7\pm 0.4$    & 53.0\\
		$C_9$ & $t_0$, $t_1$, $t_2$, $t_3$, $t_4$  & 0.063 & $-9387.3\pm 0.4$    & 103.0\\
		\hline \\
	\end{tabular}
\end{table}

\subsection{Foregrounds Modelled using Time-varying Beams}
The results shown in section \ref{s41} where the foreground signal is modelled using only one beam for multiple time-steps demonstrate that this approach can be problematic in the presence of the chromatic effects introduced by a changing ionosphere. Therefore, in this section we test time-varying beams to see how much the analyses can be improved by introducing more corresponding beams when fitting the foregrounds for cases with multiple time-steps.

Table \ref{tabmb2} shows two cases that include three time-steps where the foreground signal is modelled using two corresponding beams. We explore how the signal detection of the two time-step cases with foregrounds modelled using one constant beam would differ when introducing an extra time-step with its foregrounds modelled using its corresponding beam. This extra time-step is $t_2$, shown in Table \ref{tabmb2} in red. One observes that although $[t_0, t_1]$ as a pair does yield reasonably good evidence (Table \ref{tabpap1}), the evidence when analysed together with another time-step $t_2$ (case $C_{10}$), modelled using its corresponding beam, drops to negative values.  The other case $C_{11}$, on the other hand, as $[t_0, t_4]$ yields good evidence as a pair, yields high evidence as well as $\Delta \log\mathcal{Z}$ when analysed with the same time-step ($t_2$) as in case $C_{10}$. The RMSE also decreases which suggests better signal detection. This again shows that signal detection could be improved by including more time-steps as long as they, in smaller groups, yield better evidence and signal detection. 

\begin{table}
	\centering
	\caption{RMSE \& Log evidence $\log\mathcal{Z}$: foregrounds modelled using two beams; the foregrounds of the time-step coloured in red are modelled using its corresponding beam, and the rest using $B_{t_0}$. The cases coloured in gray are the references from Table \ref{tabmb1}. Case ($C_i$) gives the explored cases. [t] lists all the time-steps analysed in each case. RMSE is the root-mean-square error of the detected signal with respect to the injected signal. $\log\mathcal{Z}_{\mathrm{s}}$ is the evidence yielded by the foreground model including a Gaussian signal model, and $\Delta \log\mathcal{Z}$ is the difference between the evidence yielded by the foreground model including a Gaussian signal model and the one without.}
	\label{tabmb2}
	\begin{tabular}{cccrr} % n columns, alignment for each
		
		Case & [t] & RMSE (K) & $\log\mathcal{Z}_{\mathrm{s}}$ & $\Delta\log\mathcal{Z}$\\
		\hline

		$C_{10}$ & {$t_0$}, {$t_1$}, \textcolor{BrickRed}{$t_2$} & 0.056 & $-177.2 \pm 0.3$ & 9478.2\\
		\textcolor{mygray}{$C_5$} & \textcolor{mygray}{$t_0$, $t_1$, $t_2$}         & \textcolor{mygray}{0.082} & \textcolor{mygray}{$-4720.2\pm 0.4$}  & \textcolor{mygray}{194.4}\\
		$C_{11}$ & {$t_0$}, \textcolor{BrickRed}{$t_2$}, {$t_4$} & 0.022 & $796.9 \pm 0.4$ & 385.5\\
		\textcolor{mygray}{$C_6$} & \textcolor{mygray}{$t_0$, $t_2$, $t_4$}         & \textcolor{mygray}{0.034} & \textcolor{mygray}{$752.6  \pm 0.4$}    & \textcolor{mygray}{316.1}\\
		\hline \\
	\end{tabular}
\end{table}

\subsection{Deviation in Modelled Ionospheric Parameters}
\begin{figure}
    \centering
    \minipage{0.532\textwidth} \includegraphics[trim={0cm 0cm 0 0.8cm},clip,width=\linewidth]{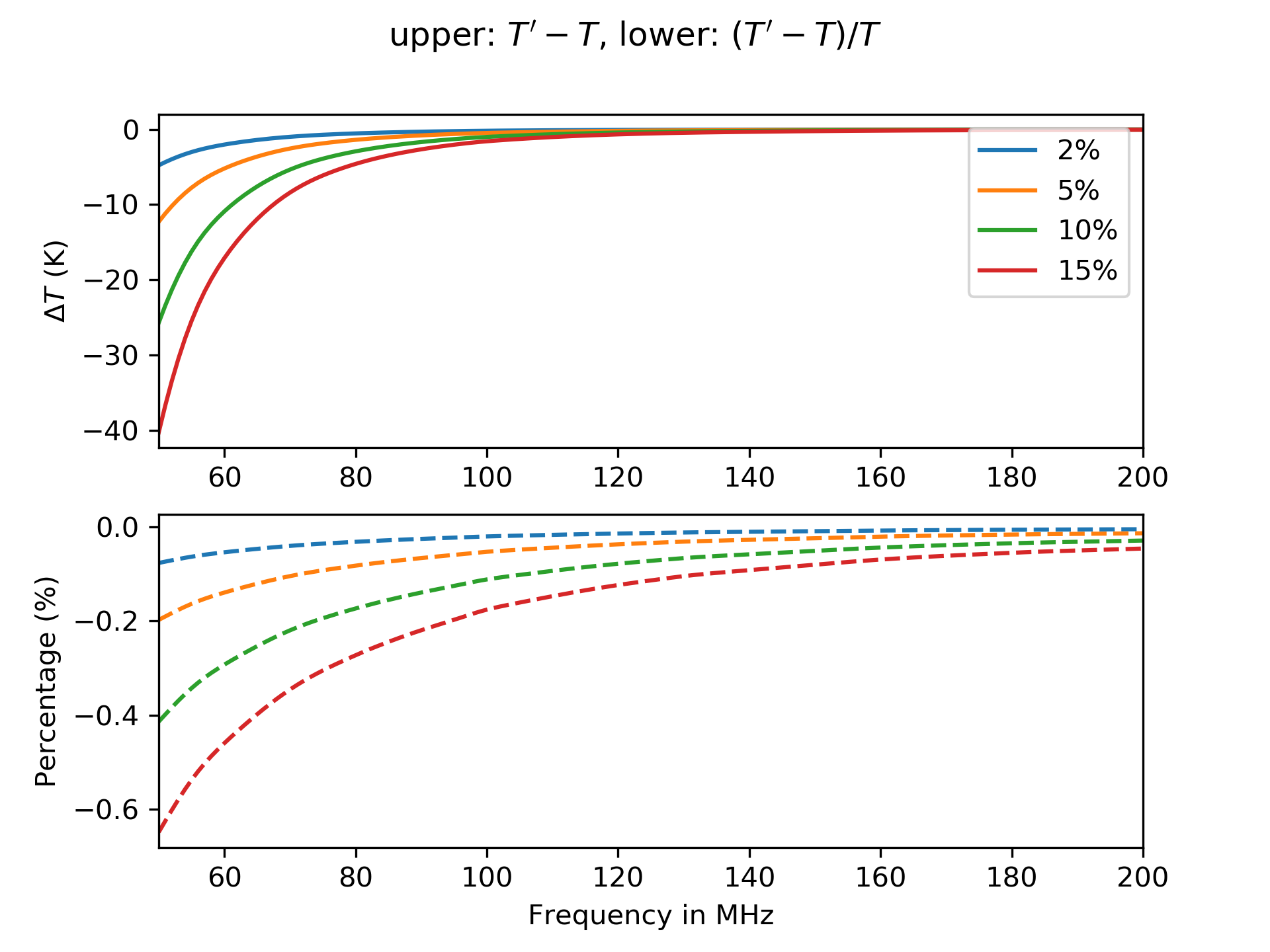}
    \endminipage\hfill
    \caption{The change in integrated antenna temperature $\Delta T = T' - T$ and the percentage of change $\Delta T/T$ with respect to frequency when a fixed amount of error [2\%, 5\%, 10\%, 15\%] is applied to the ionospheric parameters, electron density $N_\mathrm{e}$, collision frequency $\nu_\mathrm{c}$, and layer thickness, when modelling the antenna beam subjected to chromatic ionospheric effects.}
    \label{figerror}
\end{figure}

\begin{figure}
    \centering
    \minipage{0.432\textwidth} \includegraphics[trim={0cm 2cm 0 0.7cm},clip,width=\linewidth]{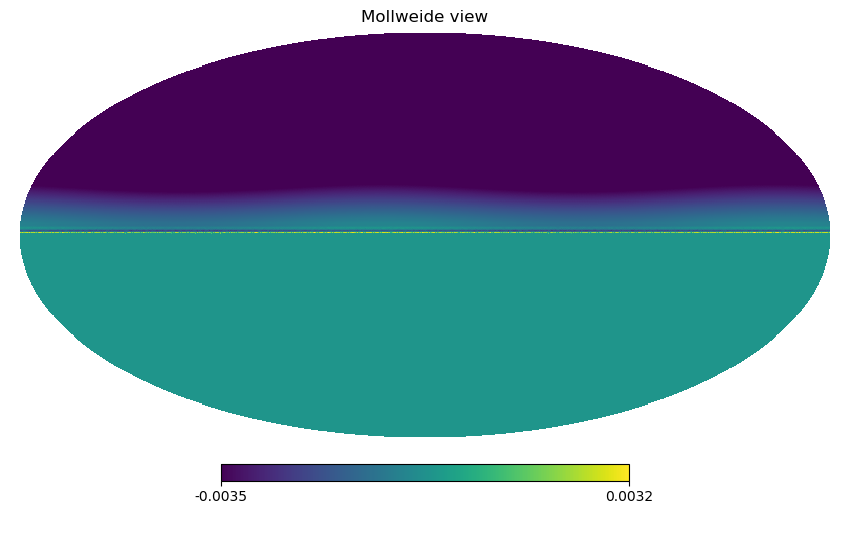}
    \endminipage\hfill
    \minipage{0.432\textwidth} \includegraphics[trim={0cm 2cm 0 0.7cm},clip,width=\linewidth]{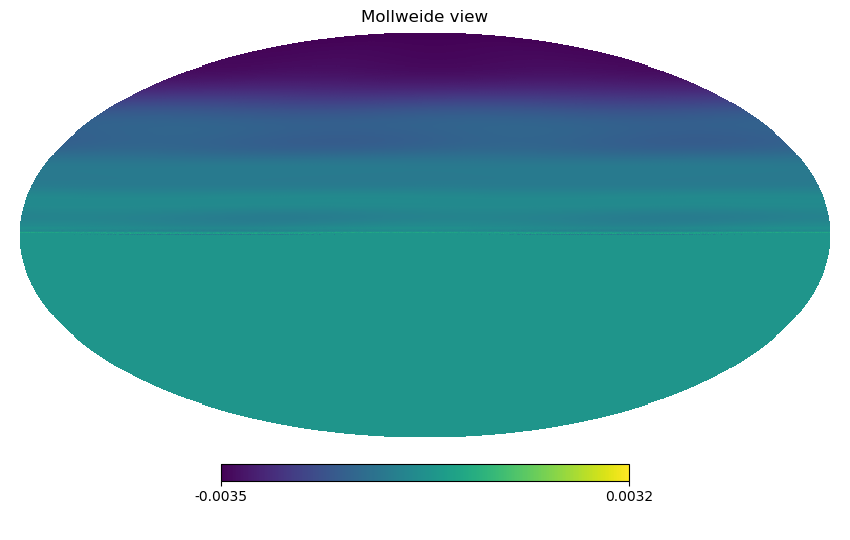}
    \endminipage\hfill
    \minipage{0.432\textwidth} \includegraphics[trim={0cm 0cm 0 0.7cm},clip,width=\linewidth]{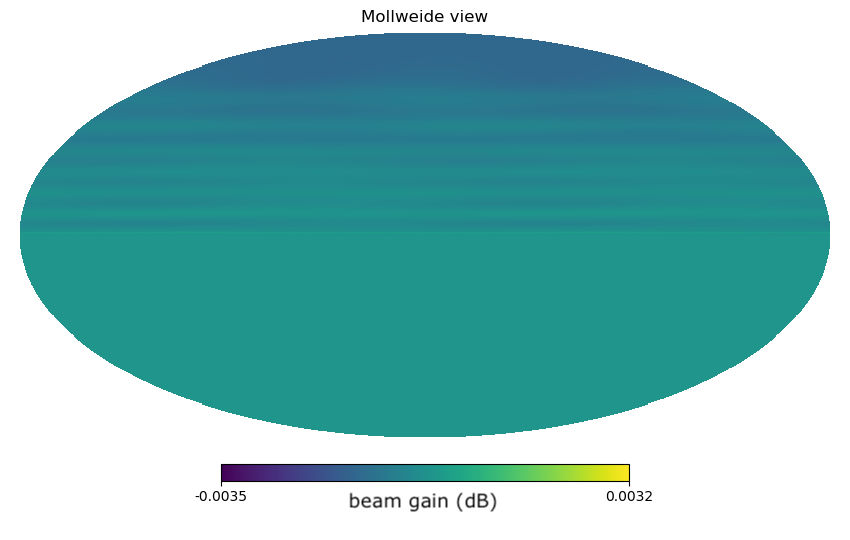}
    \endminipage\hfill
    \minipage{0.432\textwidth} \includegraphics[trim={0cm 13cm 0 0.7cm},clip,width=\linewidth]{Image/s10_50.png}
    \endminipage\hfill
    \caption{Difference in beam gain of the case modelled with 10\% deviation in ionospheric parameters with respect to the case where no deviation is applied at 50 MHz, 125 MHz, and 200 MHz, from top to bottom. The unit of the color code is dB.}
    \label{figerror10}
\end{figure}

In this section, a fixed amount of error [2\%, 5\%, 10\%, 15\%] is applied to the ionospheric parameters (a fixed percentage of increment in electron density $N_\mathrm{e}$, collision frequency $\nu_\mathrm{c}$, and layer thickness), when generating the antenna temperature data to determine how much error would lead to unsuccessful detection. A steady decrease in antenna temperature $T_A$, due to the applied error increasing the chromatic ionospheric absorption, is shown in Fig. \ref{figerror}. The data analysis pipeline then fit the foregrounds to the antenna temperature data using the beam distorted by the ionosphere, but with no error applied to the parameters. Fig. \ref{figerror10} shows how much the beam gain changes at 50, 125 and 200 MHz when a 10\% deviation (10\% increment) is applied to the ionospheric parameters with respect to when no deviation is applied.

\subsubsection{Two time-steps, foregrounds modelled using one beam and two corresponding beams}
\label{ss431}

\begin{figure*}
\centering
    \begin{tabular}{ccc}
        & foregrounds modelled with 1 beam & foregrounds modelled with $N$ beams \\
        {\rotatebox[origin=c]{90}{\hide{xxxxxxxxxxxxxxxxxxxxxxxxxxxxxxxxxxxx} \textbf{$N=2$}}} & \includegraphics[trim={0 1.25cm 0.7cm 0cm},clip,height=5.3cm]{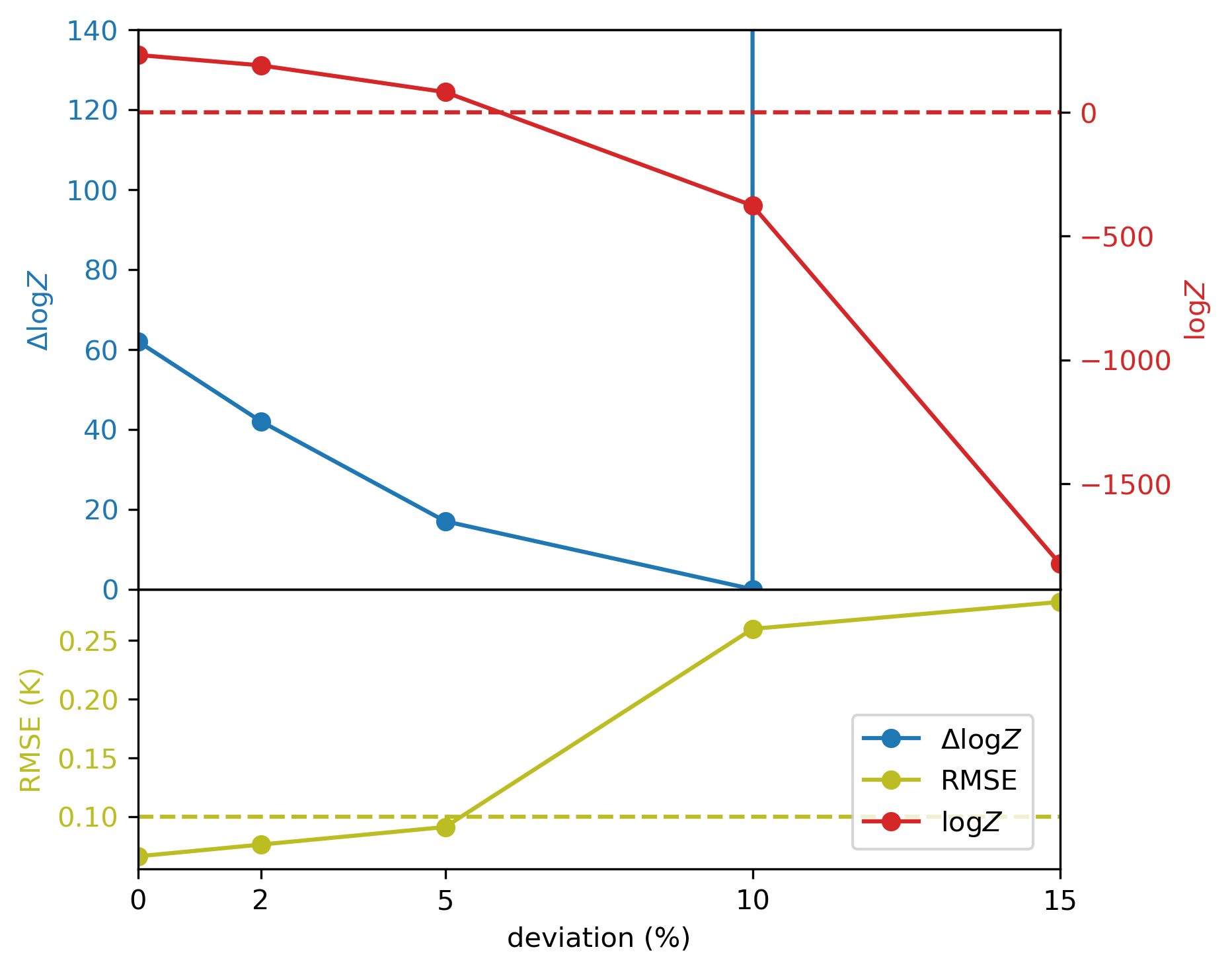} & \includegraphics[trim={0.7cm 1.25cm 0cm  0},clip,height=5.3cm]{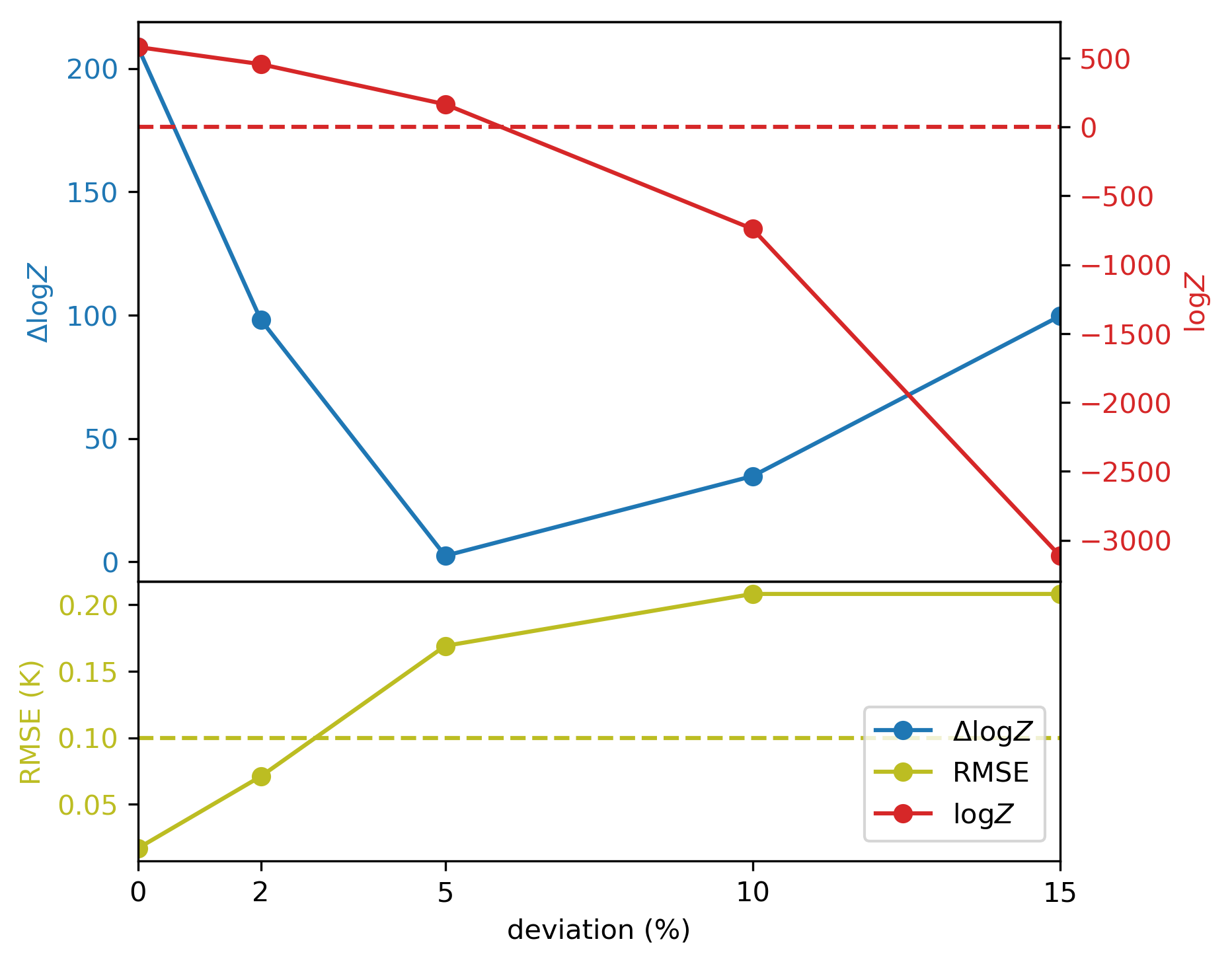}\\[-80pt]
       {\rotatebox[origin=c]{90}{\hide{xxxxxxxxxxxxxxxxxxxxxxxxxxxxxxxxxxxxxxxxxxxxx}\textbf{$N=3$}}} & \includegraphics[trim={0cm 0 0.7cm  0},clip,height=5.9cm]{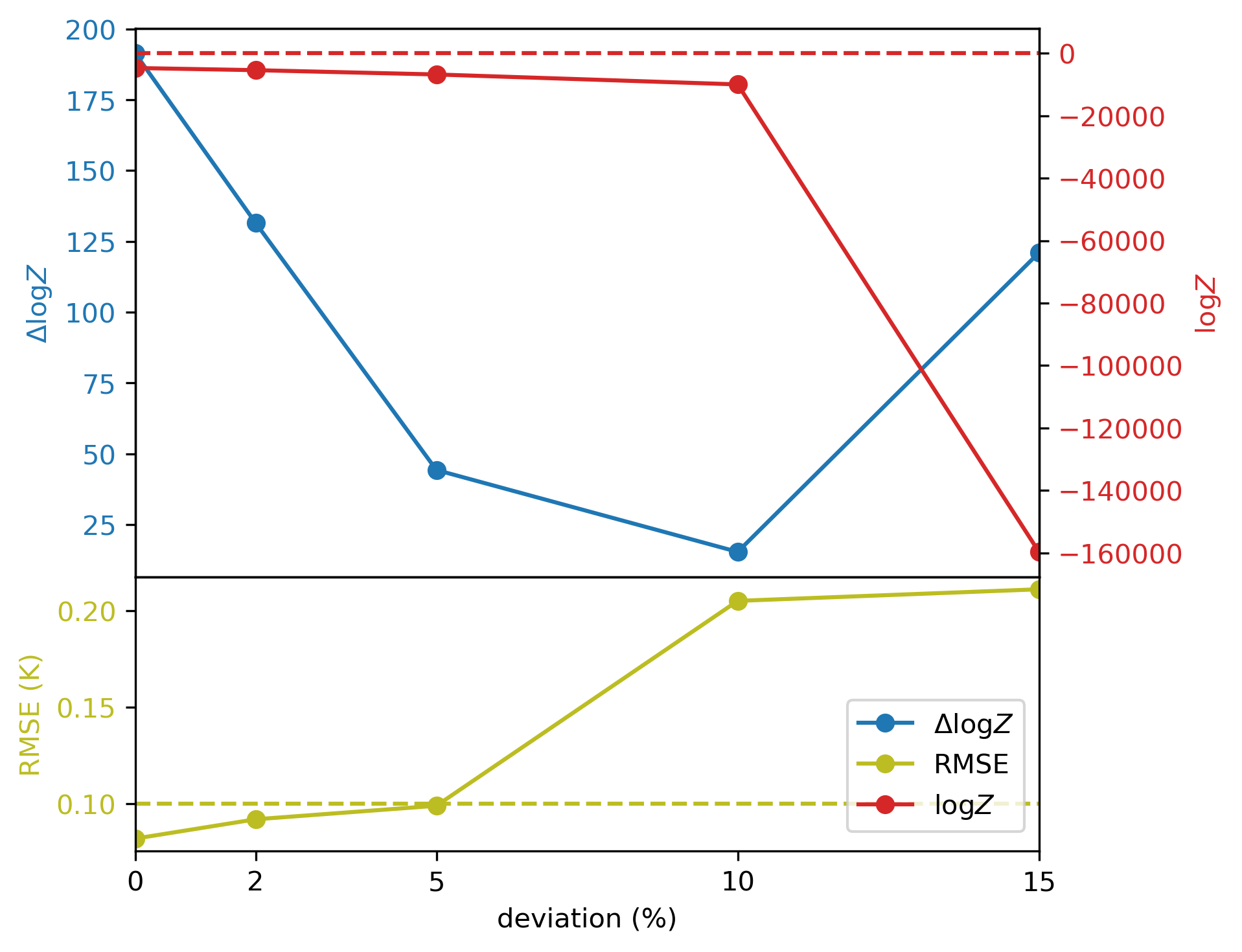} &
        \includegraphics[trim={0.7cm 0 0cm  0},clip,height=5.9cm]{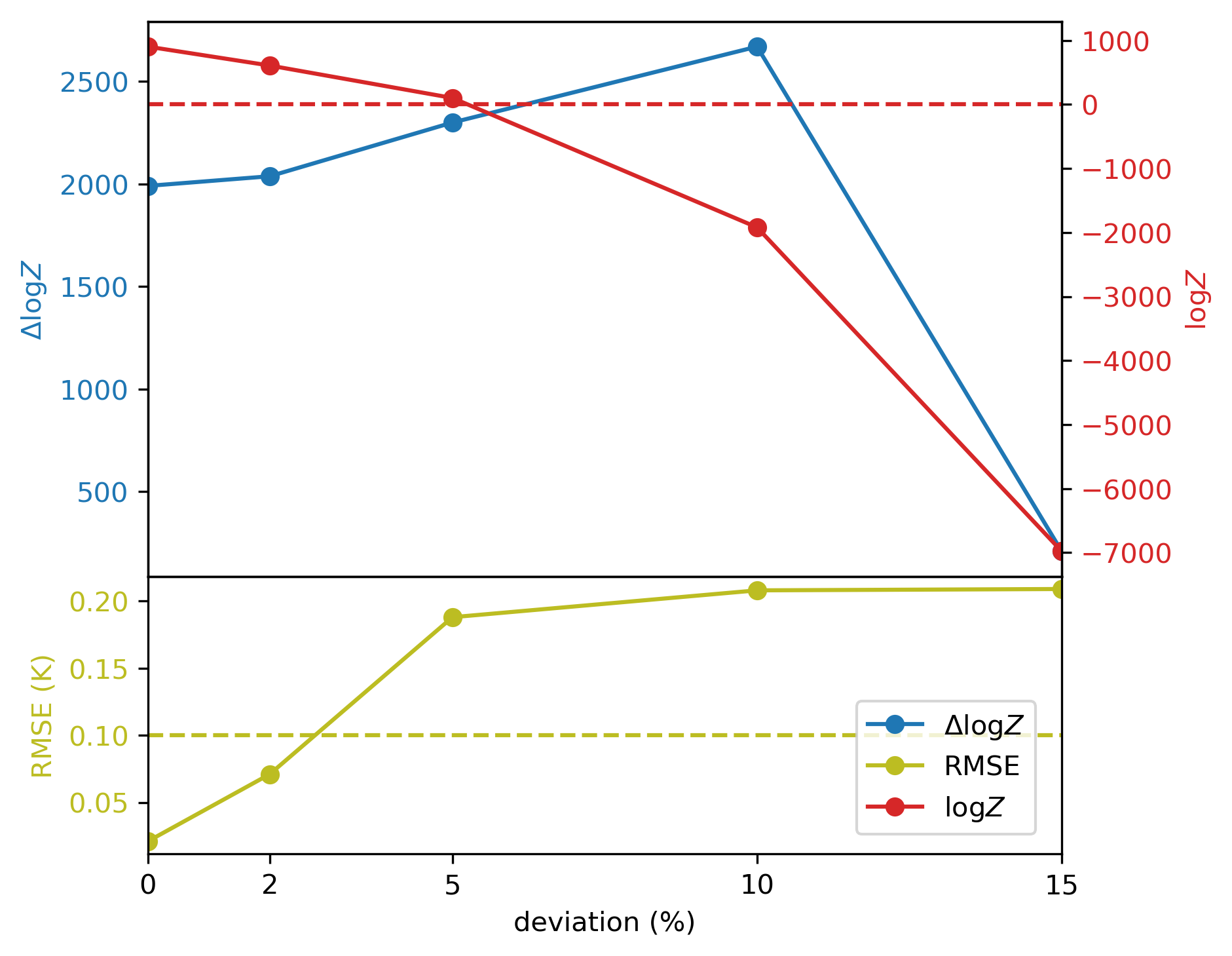}\\[-100pt]
    \end{tabular}
    \caption{$\Delta \log{\mathcal{Z}}$ (blue), $\log{\mathcal{Z}}$ (red), and RMSE (olive) of the cases where a fixed amount of error [2\%, 5\%, 10\%, 15\%] is applied to the ionospheric parameters, electron density $N_\mathrm{e}$, collision frequency $\nu_\mathrm{c}$, and layer thickness, when modelling the antenna beam subjected to chromatic ionospheric effects. The foregrounds in the data analysis pipeline are modelled using one single beam ($B_{t_0}$). The red dashed line is the 0 reference for $\log{\mathcal{Z}}$ and the olive dashed line is the 0.10K reference for RMSE. \textbf{Top-left:} two time-step ($N=2$) case [$t_0, t_1$], foregrounds modelled with one beam (Table \ref{tabpap2}). \textbf{Top-right:} two time-step ($N=2$) case [$t_0, t_1$] (Table \ref{tabpap23}), foregrounds modelled with two beams. \textbf{Bottom-left:} three time-step ($N=3$) case [$t_0, t_1, t_2$], foregrounds modelled with one beam (Table \ref{tabpap4}). \textbf{Bottom-right:} three ($N=3$) case [$t_0, t_1, t_2$], foregrounds modelled with three beams (Table \ref{tabpap3}). All plots show increasing RMSE and decreasing $\log{\mathcal{Z}}$ with deviation, which is expected. $\Delta \log{\mathcal{Z}}$  does not show a consistent trend, and sometimes even increase abruptly in the cases with  higher deviation; the value of $\Delta \log{\mathcal{Z}}$ is not reliable when the RMSE is high, as it implies the model is fitting the foregrounds instead of the injected signal. Overall, it suggests that the detection is not reliable as the deviation goes beyond 5\%.}
    
    \label{figevi}
\end{figure*}

Table \ref{tabpap2} shows the log evidences of the cases with different fixed deviations applied to the ionospheric parameters when the foregrounds of all the time-steps is modelled using only one beam $B_{t_0}$. The top-first panel of Fig. \ref{figevi} shows the RMSE and $\Delta \log{\mathcal{Z}}$ with respect to the deviation applied; one observes that the RMSE increases with the growing deviation, as expected. $\Delta \log{\mathcal{Z}}$ decreases except for the 15\% case, which might not seem reasonable; as the deviation reaches 10\%, the RMSE becomes really high and  $\log{\mathcal{Z}_s}$ turns negative, so the high $\Delta \log{\mathcal{Z}}$ could mean the model is fitting other features in the foregrounds instead of the injected global 21-cm signal, and hence its high value most likely does not suggest better signal detection. As the configuration of the analysis pipeline is identical for all the cases with deviations,  $\log{\mathcal{Z}_s}$ should be comparable; it drops consistently as the deviation in ionospheric parameters increases, and turns negative as the deviation reaches 10\%, also suggesting unreliable signal detection when the deviation goes beyond 10\%.

Table \ref{tabpap23} shows the log evidences of the cases with deviations applied to the ionospheric parameters when the foregrounds of the time-steps is modelled using two corresponding beams. Signal detection of all the cases improves comparing with the cases in which the foregrounds are modelled using only one beam, as expected.

\subsubsection{Three time-steps, foregrounds modelled using one beam and three corresponding beams}

Table \ref{tabpap3} shows the log evidences of the three time-step cases with different fixed deviations in ionospheric parameters when the foreground signal is modelled using three corresponding beams. The evidences of the lower deviation cases improve significantly with respect to the cases in which the foreground signal is modelled using only one beam.

Like section \ref{ss431}, where foregrounds are modelled using either one beam or three corresponding beams, both show reliable detection until the deviation reaches 10\%. The RMSE and $\Delta \log{\mathcal{Z}}$ with respect to the deviation applied is shown in lower panels of Fig. \ref{figevi}.

\begin{table}
	\centering
	\caption{RMSE \& Log evidence $\log\mathcal{Z}$: two time-steps [$t_0, t_1$] with deviation in ionospheric parameters, the foreground signal is modelled using one single beam ($B_{t_0}$). Case ($C_i$) gives the explored cases. [t] lists all the time-steps analysed in each case. RMSE is the root-mean-square error of the detected signal with respect to the injected signal. $\log\mathcal{Z}_{\mathrm{s}}$ is the evidence yielded by the foreground model including a Gaussian signal model, and $\Delta \log\mathcal{Z}$ is the difference between the evidence yielded by the foreground model including a Gaussian signal model and the one without.}
	\label{tabpap2}
	\begin{tabular}{ccrr} % n columns, alignment for each
		Deviation & RMSE (K) & $\log\mathcal{Z}_{\mathrm{s}}$ & $\Delta\log\mathcal{Z}$\\
		\hline

		0\%  & 0.066 & $231.5  \pm 0.3$ & 62.8 \\
		2\%  & 0.076 & $189.1  \pm 0.3$ & 42.1 \\
		5\%  & 0.091 & $81.0   \pm 0.3$ & 16.6 \\
		10\%  &  0.260 & $-377.1 \pm 0.4$ & 0.2  \\
		15\%  &  0.283 & $-1822.5\pm 0.4$ & $2.0E{+10}$\\
		\hline \\
	\end{tabular}
\end{table}

\begin{table}
	\centering
	\caption{RMSE \& Log evidence $\log\mathcal{Z}$: two time-steps [$t_0, t_1$] with deviation in ionospheric parameters, the foreground signal is modelled using two corresponding beams. Case ($C_i$) gives the explored cases. [t] lists all the time-steps analysed in each case. RMSE is the root-mean-square error of the detected signal with respect to the injected signal. $\log\mathcal{Z}_{\mathrm{s}}$ is the evidence yielded by the foreground model including a Gaussian signal model, and $\Delta \log\mathcal{Z}$ is the difference between the evidence yielded by the foreground model including a Gaussian signal model and the one without.}
	\label{tabpap23}
	\begin{tabular}{ccrr} % n columns, alignment for each
		Deviation & RMSE (K) & $\log\mathcal{Z}_{\mathrm{s}}$ & $\Delta\log\mathcal{Z}$\\
		\hline

		0\%  & 0.017 & $580.2  \pm 0.4$ & 208.7 \\
		2\%  & 0.071 & $455.7  \pm 0.4$ & 98.1 \\
		5\%  & 0.169 & $164.4  \pm 0.3$ & 2.4 \\
		10\%  &  0.208 & $-740.2 \pm 0.4$ & 34.7 \\
		15\%  &  0.208 & $-3113.0\pm 0.4$ & 99.6\\
		\hline \\
	\end{tabular}
\end{table}

\begin{table}
	\centering
	\caption{RMSE \& Log evidence $\log\mathcal{Z}$: three time-steps [$t_0, t_1, t_2$] with deviation in ionospheric parameters, the foreground signal is modelled using one single beam ($B_{t_0}$). Case ($C_i$) gives the explored cases. [t] lists all the time-steps analysed in each case. RMSE is the root-mean-square error of the detected signal with respect to the injected signal. $\log\mathcal{Z}_{\mathrm{s}}$ is the evidence yielded by the foreground model including a Gaussian signal model, and $\Delta \log\mathcal{Z}$ is the difference between the evidence yielded by the foreground model including a Gaussian signal model and the one without.}
	\label{tabpap4}
	\begin{tabular}{ccrr} % n columns, alignment for each

		Deviation & RMSE (K) & $\log\mathcal{Z}_{\mathrm{s}}$ & $\Delta\log\mathcal{Z}$\\
		\hline

		0\%  & 0.082 & $-4723.2  \pm 0.4$ & 191.4\\
		2\%  & 0.092 & $-5426.3  \pm 0.4$ & 131.5\\
		5\%  & 0.099 & $-6803.2  \pm 0.4$ & 44.3\\
		10\%  &  0.205 & $-9987.9  \pm 0.4$ & 15.4\\
		15\%  &  0.211 & $-15659.0 \pm 0.4$ & 121.9 \\
		\hline \\
	\end{tabular}
\end{table}

\begin{table}
	\centering
	\caption{RMSE \& Log evidence $\log\mathcal{Z}$: three time-steps [$t_0, t_1, t_2$] with deviation in ionospheric parameters, foregrounds modelled using three corresponding beams.}
	\label{tabpap3}
	\begin{tabular}{ccrr} % n columns, alignment for each

		Deviation & RMSE (K) & $\log\mathcal{Z}_{\mathrm{s}}$ & $\Delta\log\mathcal{Z}$\\
		\hline

		0\%  & 0.021 & $904.4   \pm 0.4$ & 1991.5\\
		2\%  & 0.071 & $609.2   \pm 0.4$ & 2038.0\\
		5\%  & 0.188 & $101.9   \pm 0.4$ & 2300.2\\
		10\%  &  0.208 & $-1919.3 \pm 0.4$ & 2669.7\\
		15\%  &  0.209 & $-6977.1 \pm 0.4$ & 210.4 \\
		\hline \\
	\end{tabular}
\end{table}

\section{Conclusions}
\label{sec5}
In this paper, we study how the time-varying chromatic ionospheric effects could affect the detection of the global 21-cm line using the existing data analysis pipeline. We explore cases including multiple time-steps during observation time to explore time variability of the ionosphere to study how the chromatic ionospheric effects on the foregrounds affects the signal detection using the REACH data analysis pipeline. We analyse and compare the different cases in terms of Bayesian log evidence and RMSE of the detected signal. We assume that the antenna beam pattern is distorted by the chromatic ionospheric effects and thus it changes with time depending on the ionospheric condition. In actual observation, however, the antenna does not change. To determine how to effectively achieve signal detection, we model the foregrounds using different number of beams to see if the pipeline can still achieve significant detection using fewer beams. We also study to what degree of error the pipeline can still manage a detection for different precision in the ionospheric parameters we use. The results can be summarised by the following points:

    (i) Signal detection does not necessarily improve when more time-steps are analysed together using the foreground model fitted with only one beam.

    (ii) In the case where the number of corresponding beams used to fit the foregrounds is lower than the number of time-steps, signal detection could improve with lower RMSE and higher evidences, when all its two time-step subgroups yield good enough signal detection in the first place. Nevertheless, for cases whose two time-step subgroups yield less significant detection, it is still possible to yield poorer yet statistically significant detection.
    
    (iii) Using more corresponding beams at different time-steps when modelling the foregrounds greatly improves signal detection.
    
    (iv) Less than 5\% error in our knowledge of the ionospheric parameters can lead to unsuccessful detection of the injected global 21-cm signal.

We find that the global signal can be robustly detected if the uncertainty in ionospheric modelling is below 5\%. In multiple time-step cases where the ionosphere does not differ too much at different time-steps, and when our knowledge of the ionospheric condition is good, signal detection can be significant. However, if the ionospheric condition differs drastically between the selected time-steps, the current data analysis pipeline might not be able to achieve statistically meaningful detection without changing the beam configuration with time.

\section*{Acknowledgements}
ES thanks Cambridge Trust and Taiwan Ministry of Education for their support. DA is supported by STFC and EdLA is supported by STFC Ernest Rutherford Fellowship. AF is supported by the Royal Society University Research Fellowship. 
%The Acknowledgements section is not numbered. Here you can thank helpful colleagues, acknowledge funding agencies, telescopes and facilities used etc. Try to keep it short.

%%%%%%%%%%%%%%%%%%%%%%%%%%%%%%%%%%%%%%%%%%%%%%%%%%
\section*{Data Availability}
Two sets of simulation data (integrated antenna temperature) for reference are available at \doi{10.5281/zenodo.4456385}. The ionospheric data  collected from Lowell GIRO Data Center are available at \url{https://ulcar.uml.edu/DIDBase/}.

%%%%%%%%%%%%%%%%%%%% REFERENCES %%%%%%%%%%%%%%%%%%

% The best way to enter references is to use BibTeX:

\bibliographystyle{mnras}
\bibliography{ref} % if your bibtex file is called example.bib

\begin{thebibliography}{}
\makeatletter
\relax
\def\mn@urlcharsother{\let\do\@makeother \do\$\do\&\do\#\do\^\do\_\do\%\do\~}
\def\mn@doi{\begingroup\mn@urlcharsother \@ifnextchar [ {\mn@doi@}
  {\mn@doi@[]}}
\def\mn@doi@[#1]#2{\def\@tempa{#1}\ifx\@tempa\@empty \href
  {http://dx.doi.org/#2} {doi:#2}\else \href {http://dx.doi.org/#2} {#1}\fi
  \endgroup}
\def\mn@eprint#1#2{\mn@eprint@#1:#2::\@nil}
\def\mn@eprint@arXiv#1{\href {http://arxiv.org/abs/#1} {{\tt arXiv:#1}}}
\def\mn@eprint@dblp#1{\href {http://dblp.uni-trier.de/rec/bibtex/#1.xml}
  {dblp:#1}}
\def\mn@eprint@#1:#2:#3:#4\@nil{\def\@tempa {#1}\def\@tempb {#2}\def\@tempc
  {#3}\ifx \@tempc \@empty \let \@tempc \@tempb \let \@tempb \@tempa \fi \ifx
  \@tempb \@empty \def\@tempb {arXiv}\fi \@ifundefined
  {mn@eprint@\@tempb}{\@tempb:\@tempc}{\expandafter \expandafter \csname
  mn@eprint@\@tempb\endcsname \expandafter{\@tempc}}}

\bibitem[\protect\citeauthoryear{Anstey, de Lera Acedo  \& Handley}{Anstey
  et~al.}{2021}]{anstey}
Anstey D.,  de Lera Acedo E.,   Handley W.,  2021, \mn@doi [Monthly Notices
  of the Royal Astronomical Society] {10.1093/mnras/stab1765}, 506, 2041

\bibitem[\protect\citeauthoryear{Anstey et~al.}{Anstey et~al.}{2022}]{domlst}
Anstey D.,  et~al., 2022, Monthly Notices of the Royal Astronomical Society

\bibitem[\protect\citeauthoryear{Bailey}{Bailey}{1948}]{bly}
Bailey D.~K.,  1948, \mn@doi [Terrestrial Magnetism and Atmospheric
  Electricity] {10.1029/TE053i001p00041}, 53, 41

\bibitem[\protect\citeauthoryear{{Bevins}, {Handley}, {Fialkov}, {de Lera
  Acedo}, {Greenhill}  \& {Price}}{{Bevins} et~al.}{2020}]{bevins}
{Bevins} H.~T.~J.,  {Handley} W.~J.,  {Fialkov} A.,  {de Lera Acedo} E.,
  {Greenhill} L.~J.,   {Price} D.~C.,  2020, arXiv e-prints, \href
  {https://ui.adsabs.harvard.edu/abs/2020arXiv200714970B} {p. arXiv:2007.14970}

\bibitem[\protect\citeauthoryear{{Bowman}, {Rogers}, {Monsalve}, {Mozdzen}  \&
  {Mahesh}}{{Bowman} et~al.}{2018}]{edges}
{Bowman} J.~D.,  {Rogers} A. E.~E.,  {Monsalve} R.~A.,  {Mozdzen} T.~J.,
  {Mahesh} N.,  2018, \mn@doi [Nature] {10.1038/nature25792}, \href
  {https://ui.adsabs.harvard.edu/abs/2018Natur.555...67B} {555, 67}

\bibitem[\protect\citeauthoryear{Chib \& Greenberg}{Chib \&
  Greenberg}{1995}]{sidd}
Chib S.,  Greenberg E.,  1995, \mn@doi [The American Statistician]
  {10.1080/00031305.1995.10476177}, 49, 327

\bibitem[\protect\citeauthoryear{{Datta}, {Bradley}, {Burns}, {Harker},
  {Komjathy}  \& {Lazio}}{{Datta} et~al.}{2016}]{burns}
{Datta} A.,  {Bradley} R.,  {Burns} J.~O.,  {Harker} G.,  {Komjathy} A.,
  {Lazio} T. J.~W.,  2016, \mn@doi [Astrophysical Journal]
  {10.3847/0004-637X/831/1/6}, \href
  {https://ui.adsabs.harvard.edu/abs/2016ApJ...831....6D} {831, 6}

\bibitem[\protect\citeauthoryear{{Evans} \& {Hagfors}}{{Evans} \&
  {Hagfors}}{1968}]{evans1968}
{Evans} J.~V.,  {Hagfors} T.,  1968, {Radar astronomy}.
New York McGraw-Hill

\bibitem[\protect\citeauthoryear{Foreman-Mackey, Hogg, Lang  \&
  Goodman}{Foreman-Mackey et~al.}{2013}]{Mackey2013}
Foreman-Mackey D.,  Hogg D.~W.,  Lang D.,   Goodman J.,  2013, \mn@doi
  [Publications of the Astronomical Society of the Pacific] {10.1086/670067},
  125, 306

\bibitem[\protect\citeauthoryear{Furlanetto}{Furlanetto}{2016}]{furl}
Furlanetto S.~R.,  2016, The 21-cm Line as a Probe of Reionization.
Astrophysics and Space Science Library, p. 247–280, \url
  {https://doi.org/10.1007/978-3-319-21957-8}

\bibitem[\protect\citeauthoryear{Handley, Hobson  \& Lasenby}{Handley
  et~al.}{2015}]{hand5b}
Handley W.~J.,  Hobson M.~P.,   Lasenby A.~N.,  2015, \mn@doi [Monthly Notices
  of the Royal Astronomical Society] {10.1093/mnras/stv1911}, 453, 4384

\bibitem[\protect\citeauthoryear{{Hills}, {Kulkarni}, {Meerburg}  \&
  {Puchwein}}{{Hills} et~al.}{2018}]{hillsnat}
{Hills} R.,  {Kulkarni} G.,  {Meerburg} P.~D.,   {Puchwein} E.,  2018, \mn@doi
  [Nature] {10.1038/s41586-018-0796-5}, \href
  {https://ui.adsabs.harvard.edu/abs/2018Natur.564E..32H} {564, E32}

\bibitem[\protect\citeauthoryear{Mitra}{Mitra}{1959}]{theight}
Mitra A.~P.,  1959, \mn@doi [Journal of Geophysical Research (1896-1977)]
  {10.1029/JZ064i007p00733}, 64, 733

\bibitem[\protect\citeauthoryear{{Nicolet}}{{Nicolet}}{1953}]{nicole}
{Nicolet} M.,  1953, \mn@doi [Journal of Atmospheric and Terrestrial Physics]
  {10.1016/0021-9169(53)90110-X}, \href
  {https://ui.adsabs.harvard.edu/abs/1953JATP....3..200N} {3, 200}

\bibitem[\protect\citeauthoryear{{Reis}, {Fialkov}  \& {Barkana}}{{Reis}
  et~al.}{2020}]{Reis2020}
{Reis} I.,  {Fialkov} A.,   {Barkana} R.,  2020, \mn@doi [\mnras]
  {10.1093/mnras/staa3091}, \href
  {https://ui.adsabs.harvard.edu/abs/2020MNRAS.499.5993R} {499, 5993}

\bibitem[\protect\citeauthoryear{{Reis}, {Fialkov}  \& {Barkana}}{{Reis}
  et~al.}{2021}]{Reis20211}
{Reis} I.,  {Fialkov} A.,   {Barkana} R.,  2021, \mn@doi [\mnras]
  {10.1093/mnras/stab2089}, \href
  {https://ui.adsabs.harvard.edu/abs/2021MNRAS.506.5479R} {506, 5479}

\bibitem[\protect\citeauthoryear{Shen, Anstey, de Lera Acedo, Fialkov  \&
  Handley}{Shen et~al.}{2021}]{shen21}
Shen E.,  Anstey D.,  de Lera Acedo E.,  Fialkov A.,   Handley W.,  2021,
  \mn@doi [Monthly Notices of the Royal Astronomical Society]
  {10.1093/mnras/stab429}, 503, 344

\bibitem[\protect\citeauthoryear{{Sims} \& {Pober}}{{Sims} \&
  {Pober}}{2020}]{sims}
{Sims} P.~H.,  {Pober} J.~C.,  2020, \mn@doi [Monthly Notices of the Royal
  Astronomical Society] {10.1093/mnras/stz3388}, \href
  {https://ui.adsabs.harvard.edu/abs/2020MNRAS.492...22S} {492, 22}

\bibitem[\protect\citeauthoryear{{Singh} \& {Subrahmanyan}}{{Singh} \&
  {Subrahmanyan}}{2019}]{singh}
{Singh} S.,  {Subrahmanyan} R.,  2019, \mn@doi [\apj]
  {10.3847/1538-4357/ab2879}, \href
  {https://ui.adsabs.harvard.edu/abs/2019ApJ...880...26S} {880, 26}

\bibitem[\protect\citeauthoryear{{Singh} et~al.,}{{Singh}
  et~al.}{2021}]{saras3}
{Singh} S.,  et~al., 2021, arXiv e-prints, \href
  {https://ui.adsabs.harvard.edu/abs/2021arXiv211206778S} {p. arXiv:2112.06778}

\bibitem[\protect\citeauthoryear{{Vedantham}, {Koopmans}, {de Bruyn},
  {Wijnholds}, {Ciardi}  \& {Brentjens}}{{Vedantham} et~al.}{2014}]{km}
{Vedantham} H.~K.,  {Koopmans} L.~V.~E.,  {de Bruyn} A.~G.,  {Wijnholds} S.~J.,
   {Ciardi} B.,   {Brentjens} M.~A.,  2014, \mn@doi [Monthly Notices of the
  Royal Astronomical Society] {10.1093/mnras/stt1878}, \href
  {https://ui.adsabs.harvard.edu/abs/2014MNRAS.437.1056V} {437, 1056}

\bibitem[\protect\citeauthoryear{de Lera~Acedo}{de~Lera~Acedo}{2019}]{reachh}
de Lera~Acedo E.,  2019, \mn@doi [2019 International Conference on
  Electromagnetics in Advanced Applications (ICEAA)]
  {10.1109/ICEAA.2019.8879199}, pp 0626--0629

\makeatother
\end{thebibliography}

% Alternatively you could enter them by hand, like this:
% This method is tedious and prone to error if you have lots of references
%\begin{thebibliography}{99}
%\bibitem[\protect\citeauthoryear{Author}{2012}]{Author2012}
%Author A.~N., 2013, Journal of Improbable Astronomy, 1, 1
%\bibitem[\protect\citeauthoryear{Others}{2013}]{Others2013}
%Others S., 2012, Journal of Interesting Stuff, 17, 198
%\end{thebibliography}

%%%%%%%%%%%%%%%%%%%%%%%%%%%%%%%%%%%%%%%%%%%%%%%%%%

%%%%%%%%%%%%%%%%% APPENDICES %%%%%%%%%%%%%%%%%%%%%

\iffalse
\appendix

\section{hello}
\fi

%%%%%%%%%%%%%%%%%%%%%%%%%%%%%%%%%%%%%%%%%%%%%%%%%%

% Don't change these lines
\bsp	% typesetting comment
\label{lastpage}
\end{document}